\documentclass[aps,pra,showpacs,onecolumn]{revtex4}

\usepackage{graphicx}

\usepackage{amssymb}
\usepackage{amsmath}
\usepackage{braket}
\usepackage{xcolor}
\usepackage{hyperref}
\usepackage{color}
\usepackage{pbox}
\usepackage{float}
\usepackage{enumerate}

\newcounter{nota}
\newcommand{\nota}[1]{ \stepcounter{nota}
                 \noindent{\bf Nota \Roman{nota}: }  }

\newtheorem{teo}{Theorem}

\begin{document}

\title{Exact solution of a family of staggered Heisenberg chains with conclusive
pretty good quantum state transfer}

\author{Pablo Serra$^{(1)}$, Alejandro Ferr\'on$^{(2)}$, and
Omar Osenda$^{(1)}$}

\affiliation{
(1) Instituto de F\'sica Enrique Gaviola (CONICET-UNC) and Facultad de 
Matem\'atica, Astronom\'ia, F\'isica y Computaci\'on, Universidad Nacional de 
C\'ordoba,
Av. Medina Allende s/n, Ciudad Universitaria, CP:X5000HUA C\'ordoba, Argentina\\
(2) Instituto de Modelado e Innovaci\'on Tecnol\'ogica
(CONICET-UNNE) and
Facultad de Ciencias Exactas, Naturales y Agrimensura, Universidad Nacional
del Nordeste, Avenida Libertad 5400, W3404AAS Corrientes, Argentina.
}

\date{\today}

\begin{abstract}
We construct the exact solution for a family of one-half spin chains explicitly. 
The spin chains
Hamiltonian corresponds to an isotropic Heisenberg Hamiltonian, with staggered 
exchange couplings
that take only two different values. We work out the exact solutions in the one-
excitation subspace. Regarding the problem of quantum state transfer, we use the 
solution
and some theorems concerning the approximation of irrational numbers, to show 
the appearance
of conclusive pretty good transmission for chains with particular lengths. We 
present numerical evidence that pretty good transmission is achieved by chains 
whose length is not a power of two. The set of spin chains that shows pretty 
good transmission is a subset of
the family with an exact solution. Using perturbation
theory, we thoroughly analyze the case when one of the exchange coupling 
strengths is orders of magnitude larger than
the other.  This strong coupling limit allows us to study, in a simple way, the 
appearance of pretty
good transmission. The use of analytical closed expressions for the eigenvalues, 
eigenvectors, and
transmission probabilities allows us to obtain the precise asymptotic behavior 
of the time where the
pretty good transmission is observed. Moreover, we show that this time scales as 
a power
law whose exponent is an increasing function of the chain length. We also 
discuss the crossover
behavior obtained for the pretty good transmission time between the regimes of 
strong coupling
limit and the one observed when the exchange couplings are of the same order of 
magnitude.

\end{abstract}

\maketitle

\section{introduction}

The Heisenberg model has been studied in the context of the Quantum state 
Transfer in a host of works \cite{Bose2003,Bose-review,Nikolopoulos2015}, with 
different strategies that go from rigorous 
results and theorems \cite{Kay2019,Kay2010,Banchi2017,vanBommel2010}to 
completely numerical efforts. Even the availability of 
the Bethe ansatz, and the integrability of the model, can not hide the fact 
that 
the autonomous evolution of homogeneous Heisenberg chains is ill-suited to 
perform Quantum State Transfer (QST) with very high fidelity and short transfer 
times. The XX chains, on the contrary, allow perfect quantum state transfer 
using different strategies \cite{Zwick2011,Mograby2021,VZ2017,VZ2} or near 
perfect QST at transfer times close to the 
minimal time allowed by the Quantum Speed limit 
\cite{Zwick3,Banchi2010,Banchi2011}. Regrettably, more and more 
experimental implementations can be modeled only by using effective Heisenberg 
chain Hamiltonians  
\cite{Kandel2021,Martins2017,Kostak2007,quantum-dot-chain,Li2018,
nuclear-spin-chain,Loft2011,Banchi2011prl,Chapman2016,Kandel2019,Baum2021} (or 
more complicated interactions \cite{Hauke2010,Porras2004,Hung2016})  while the 
XX chains 
remain as an option only for simplistic highly anisotropic situations. Of 
course, there is always room for renewed proposals that involve the XX 
Hamiltonian, as is the case in some XX chains that show topological states 
\cite{Hauke2010,Porras2004,Hung2016}.

On the other hand, much progress have been made concerning the non-autonomous 
time evolution of spin chains 
\cite{Wang2016,Burgarth2010,Yang2010,Zhang2016,Farooq2015,Coden2021}, although 
many times the control leads to particular results about the times for which 
the transfer is achievable 
\cite{Wang2016,Burgarth2010,Yang2010,Zhang2016,Farooq2015}. The controllability 
of the dynamics is a well studied area with numerous results 
\cite{Jurdjevic1972,Burgarth2009,Wang2016,Ramakrishna1995}.

The basic difficulty of the homogeneous Heisenberg chains mentioned above has 
driven  the study of some alternatives and the formulation of a 
specialized jargon that classifies the different dynamical scenarios that arise 
when non-homogeneous chains are considered or the condition of a short (and 
known) arrival time is lifted \cite{Kay2019,Kay2010,Banchi2017,vanBommel2010}. 
The pretty good QST \cite{Banchi2017,Kay2019} and fractional revival 
\cite{Vinet1,Vinet2} are 
two of the most studied scenarios, the spectral conditions needed to their 
appearance are well understood. Nevertheless, these conditions do not 
prescribe, 
for instance, how the different exchange coefficients must be selected to 
achieve the pretty good scenario nor the precise arrival time at which the QST 
is effectively observed or how this time could scale with the two main 
parameters of the chain, the largest interaction strength, and its length.

Heisenberg spin chains with nearest-neighbor time-independent staggered, or 
alternating, exchange coefficients are easier to implement than 
chains 
with all their spin coupling interactions tuned to precise values, at least in 
principle. Despite that 
the tailoring of all couplings can be, in principle,  achieved in some 
implementations and result in near-perfect QST at known times 
\cite{Serra2022,Ferron2022}, in this paper we 
aim to study the simpler staggered case, when the exchange couplings can take 
only two alternating values, $J_1$ and $J_2$. The typical chain has an exchange 
coupling (EC) equal to $J_1$ between the first and second spin, $J_2$ between 
the second and third spin, and so on. The sequence described results in a 
centrosymmetric chain if the number of spins is even, with the central EC equal 
to $J_2$. 

First, we will show that for particular families of chain lengths the spectrum 
and eigenvalues of the problem restricted to a subspace of the whole Hilbert 
space can be calculated exactly using rather simple means. More importantly, 
the 
functional dependency of the eigenvalues with the parameters allows for a 
detailed analysis of the properties of the QST process through the population 
transfer. Using theorems about the best rational approximation that can be  
achieved for irrational numbers we can explicitly show the appearance of the 
pretty good QST scenario for small chains. Besides we can calculate explicitly 
the arrival time, {\em i.e.} we obtain the time $t_{\epsilon}$ such that the 
population transferred satisfies $P(t_{\epsilon}) < 1-\epsilon$. Interestingly, 
for small chains, it is possible to show that the scenario appears for almost 
any 
value of the parameter $J_2$ while keeping $J_=1$ and for chains with $N=2^k$ 
and $N=\alpha\times 
2^k $, with $\alpha=3$ or $5$. This last result shows that the family of spin 
chains possessing PGT is 
larger than previously reported.

Later, we present analytical results for the population transferred in the 
limit $J_2\longrightarrow \infty$. We call this regime the strong coupling 
limit. We work out the results using perturbation theory and show that the 
population 
transferred in the strong coupling limit has a simpler structure and dependency 
with the eigenvalues than the 
exact results. The strong coupling limit allows us to study chains of arbitrary 
length. We present evidence that the 
population in the strong coupling limit also shows pretty good QST. Together 
with the results found for 
small chains, we  conjecture the scaling of $t_{\epsilon}\sim 
1/(\epsilon)^{f(N)}$, 
where $f(N)$ is a simple function of the chain length.

\section{The Quantum State Transfer protocol and the transferred 
population}\label{sec:model}

We  focus our study on the well-known Heisenberg quantum spin chain 
Hamiltonian, 

\begin{equation}
 H = - \sum_{i=1}^{N-1} J_i\,\vec{\sigma}_i \cdot \vec{\sigma}_{i+1}
	\label{eh1}
\end{equation}

\noindent where $J_i>0$. The spectral and transfer properties of this 
Hamiltonian have been extensively studied, in particular  when the chains are 
homogeneous, 
{\em i.e.} all the $J_i$ coefficients  in Eq.~\eqref{eh1} are equal ($J_i=J, 
\forall i$). The homogeneous chain is quite poor as a 
transfer channel when  the simplest transfer protocol is used.

The Hamiltonian 
in Eq. (\ref{eh1}) commute with the total magnetization in the $z-$direction 

\begin{equation}
 \left[ H, \sum_i \sigma_i^z \right] = 0,
\end{equation}

\noindent so it can be diagonalized in subspaces with 
fixed number of excitations, {\em i.e.} in subspaces with a given number of 
spins up. It is customary to use the computational basis, where for a single 
spin $|0\rangle =|\downarrow\rangle$ and $|1\rangle =|\uparrow\rangle$, so
$|\mathbf{0}\rangle =|0000\ldots 0\rangle$ is the state with zero spins up of 
the whole chain. 

Throughout this paper, we will work in the single excitation Hilbert space and 
study $h$, the $N\times N$ matrix of this block. We are mainly interested in 
finding the exact eigenvalues and eigenvectors of $h$ and using them to study 
transmission in this kind of spin chain. We understand that an exactly-solvable 
spin chain is one such that its eigenvalues and eigenvectors have an analytic 
expression. Once we have the eigenvalues and eigenvectors we should analyze the 
transmission properties of our spin chain. So we introduce a quantity that 
describes the transmission probability 
between an initial state and a final state. The $N$ states with a single spin up 
are denoted as follows

\begin{equation}
|\mathbf{1} \rangle = |10\ldots 0\rangle, |\mathbf{2} \rangle =
|010\ldots0\rangle,\ldots , |\mathbf{N} \rangle = |00\ldots 1\rangle.
\end{equation}

\noindent {\em i.e.} the state $|\mathbf{j} \rangle $ is the state of the chain
with only the $j$-th spin up. In the protocol studied along this paper the 
initial state of the chain, $|\Psi(0)\rangle$, is prepared as

\begin{equation}\label{eq:initial-chain-product}
|\Psi(0)\rangle =|\psi(0)\rangle \otimes |\mathbf{0}\rangle_{N-1} ,
\end{equation}

\noindent where $|\psi(0)\rangle=|1\rangle$ is an excitation in the first 
site of the chain and $|\mathbf{0}\rangle_{N-1}$ is the state
without excitations of a chain with $N-1$ spins. The state in
Eq.~\eqref{eq:initial-chain-product} can be rewritten as

\begin{equation}
 \label{eq:initial-state}
|\Psi(0)\rangle =|\mathbf{1} \rangle .
\end{equation}
Using the time evolution operator
\begin{equation}
\label{eq:time-evolution}
U(t) = \exp{(-i H t)},
\end{equation}
the state of the chain at time $t$ can be obtained as
\begin{equation}
 \label{eq:time-dependent-state}
|\Psi(t)\rangle = U(t) |\Psi(0)\rangle = U(t) |\mathbf{1}\rangle .
\end{equation}
\noindent and we define the transferred population (TP) between the first and 
last 
sites of the chain as

\begin{equation}\label{eq:population}
P(t) = |\langle \mathbf{1}|U(t)|\mathbf{N}\rangle|^2 ,
\end{equation}

\noindent this is our quantity of interest. Perfect transmission (PT) is 
achieved when $P(t)=1$ for some time.  The scenario known as 
pretty good quantum state transfer (PGT) occurs if for some time $t_{\epsilon}$

\begin{equation}
 P(t_{\epsilon}) = 1- \epsilon, \quad \forall \epsilon>0 .
\end{equation}

\section{Exact solution for spin chains with staggered exchange coupling 
coefficients}

In this paper, we focus on non-homogeneous spin chains    
obtaining analytical and closed expressions for eigenvalues, eigenvectors, and 
transferred population or transmission probabilities. It is important to note 
that it is always possible to take $J_k=1$  for some $k$, so chains with $N=2$  
have no free parameters, and chains with $N=3$ have just one free parameter 
($J_1=1; J_2$ free). It is easy to show that, for $N=4$, there is no 
analytically closed solution when $J_1=1$ and $J_2,\;J_3$ are arbitrary 
parameters. Obtaining analytical expressions in a completely general scenario 
would be impossible. We start asking for  centrosymmetric spin chains as in most 
of our previous works \cite{Coden2021,Serra2022,Ferron2022}

\begin{equation}
 \label{eccs}
J_{N-i}=J_i;\; i=1,\cdots,[N/2] \,.
\end{equation}

\noindent  With this convention, the matrix  $h$ results bisymmetric, and the 
spin chains with $N=4,\;5$  are now exactly-solvable. Anyway, $N  \geqslant 6$ 
has no analytical expressions for its eigenvalues and eigenvectors. We will show 
that the bisymmetric spin chains are exactly-solvable when their length is~
$N=\alpha \times 2^k$, with $\alpha$ and $k$ natural numbers . We 
choose  centrosymmetric  chains with

\begin{equation}
 \label{eces}
J_{2 i-1}=1 \quad\; \quad J_{2 i}=J_2 \quad;\quad i=1,\cdots,N/2 
\,;\quad N \in  2\mathbb{N} \,.
\end{equation}

\noindent In this case, the corresponding matrix $h^{(N)}$ can be written as:


\begin{equation}
\label{emhg}
h^{(N)}\,=\,
\begin{pmatrix}
d^{(N)}&  -2&  0 & 0&0&0& \ldots &  0 &  0&0 \\
-2 & d^{(N)}+2\,J_2 &  -2\,J_2& 0&0&0& \ldots &  0 &  0 &0\\
0 &  -2\,J_2&   d^{(N)}+2\,J_2 &  -2&  0 &  0 & \ldots &  0 &  0&0 \\
0 &  0 &  -2 &   d^{(N)}+2\,J_2 & -2\,J_2&  0& \ldots &  0 &  0&0 \\
0 &  0&0&-2\,J_2&  d^{(N)}+2\,J_2 &  -2 & \ldots &  0 &  0 &0\\
\vdots&\vdots&\vdots&\vdots&\vdots&\vdots&\ddots&\vdots&\vdots\\
0 &  0&0&0&  0 &  0 & \ldots &   d^{(N)}+2\,J_2 &  -2\,J_2&0  \\
0 &  0&0&0&  0 &  0 & \ldots &  -2\,J_2 &   d^{(N)}+2\,J_2&-2  \\
0 &  0 & 0&0& 0& 0 &\ldots &  0& -2& d^{(N)}
\end{pmatrix} \,,
\end{equation}


\noindent where

\begin{equation}
\label{edn}
d^{(N)}=-\left(\frac{N}{2}-2\right) - \left(\frac{N}{2}-1\right)\;J_2 \,.
\end{equation}

It is convenient to define the matrix

\begin{equation}
\label{enmch}
\mathfrak{h}^{(N)}\,=\,h^{(N)}-(d^{(N)}+2\,J_2)\,\mathbb{I}_N \,,
\end{equation}

\noindent because its elements do not depend on $N$. Clearly, if $\lambda$ is 
an eigenvalue of $\mathfrak{h}^{(N)}$, then $\lambda+d^{(N)}+2\,J_2$ is
an eigenvalue of $h^{(N)}$, and both matrices have
the same eigenvectors. $h^{(N)}$, and
then $\mathfrak{h}^{(N)}$, are   bisymmetric matrices, then  we can  write:

\begin{equation}
\label{emhrbs}
\mathfrak{h}^{(N)}\,=\,
\begin{pmatrix}
A & B  \\
B^t& S\,A\,S
\end{pmatrix} \,,
\end{equation}

\noindent where the $N/2 \times N/2$ matrices $A,\,B$ and $S$ are

\begin{equation}
\label{emarg}
A\,=\,
\begin{pmatrix}
-2\,J_2&  -2&  0 & 0&0&0& \ldots &  0 &  0&0 \\
-2 & 0 &  -2\,J_2& 0&0&0& \ldots &  0 &  0 &0\\
0 &  -2\,J_2&   0 &  -2&  0 &  0 & \ldots &  0 &  0&0 \\
0 &  0 &  -2 &   0 & -2\,J_2&  0& \ldots &  0 &  0&0 \\
0 &  0&0&-2\,J_2&  0 &  -2 & \ldots &  0 &  0 &0\\
\vdots&\vdots&\vdots&\vdots&\vdots&\vdots&\ddots&\vdots&\vdots\\
0 &  0&0&0&  0 &  0 & \ldots &   0 &  b&0  \\
0 &  0&0&0&  0 &  0 & \ldots &  b &   0&a  \\
0 &  0 & 0&0& 0& 0 &\ldots &  0& a& 0
\end{pmatrix} \,;
\end{equation}

\begin{equation}
\label{embrg}
B\,=\,\begin{pmatrix}
  0&  0 &0& \ldots &  0 &  0 \\
  0&  0 &0& \ldots &  0 &  0 \\
\vdots&\vdots&\vdots&\ddots&\vdots&\vdots\\
0 &  0&0& \ldots &   0 &0  \\
b &  0&0& \ldots &   0&0  \\
\end{pmatrix} \quad;\quad
S\,=\,\begin{pmatrix}
  0&  0 &0& \ldots &  0 &  1 \\
  0&  0 &0& \ldots &  1 &  0 \\
\vdots&\vdots&\vdots&\ddots&\vdots&\vdots\\
0 &  1&0& \ldots &   0 &0  \\
1 &  0&0& \ldots &   0&0  \\
\end{pmatrix}  \,,
\end{equation}

\noindent where $a$ and $b$ are constants depending on the parity of $N/2$,

\begin{equation}
\label{epayb}
a\,=\,\left\{ \begin{array}{ll}
-2 &\mbox{if  $N/2$ is even } \\
-2\,J_2 &\mbox{if  $N/2$ is odd }
\end{array} \right.  \quad;\quad
b\,=\,\left\{ \begin{array}{ll}
-2\,J_2 &\mbox{if  $N/2$ is even } \\
-2 &\mbox{if  $N/2$ is odd }
\end{array} \right.  \;.
\end{equation}

\noindent Now, defining $A_\pm^{(N/2)}=A\,\pm \,B\,S$ we know
that the eigenvalues of $A_-^{(N/2)}$ and $A_+^{(N/2)}$ are the
eigenvalues of $\mathfrak{h}^{(N)}$ \cite{nouri2012}. In the 
particular case where $N/2$ is even, we have  the important relation

\begin{equation}
A_+^{(N/2)}\,=\,\mathfrak{h}^{(N/2)} \label{ean2} \,,
\end{equation}

\noindent that is, if we know the eigenvalues of $\mathfrak{h}^{(N/2)}$, to know
the eigenvalues of $\mathfrak{h}^{(N)}$ we only need to calculate the 
eigenvalues of $A_-^{(N/2)}$. By inspection, we corroborate that, for even
 values of $N$,
 the eigenvalues of $\mathfrak{h}^{(N)}$ can be written as 
 
\begin{equation}
\label{eavptnp}
\begin{split}
\lambda_-^{(m)} &=
- 2\,\sqrt{1+J_2^2 \,+\, 2\,J_2\cos\left(\frac{2\,m\,\pi}{N}\right)}
\;;\;m=0,\ldots,\,\frac{N}{2}\\ 
\lambda_+^{(m)} &=\;
 \,2\,\sqrt{1+J_2^2 \,+\, 2\,J_2\cos\left(\frac{2\,m\,\pi}{N}\right)}
\;;\;m=1,\ldots,\,\frac{N}{2}-1
\,.
\end{split}
\end{equation}

For the case $N/2$ even, the eigenvalues of $A_-^{(N/2)}$ are given by
the odd values of $m$,

\begin{equation}
\label{eavptnpam}
\lambda_\pm^{(2l+1)} \,=\,
\pm 2\,\sqrt{1+J_2^2 \,+\, 2\,J_2\cos\left(\frac{(2\,l+1)\,\pi}{(N/2)}\right)}
\;;\;l=0,\ldots,\,\frac{N}{4}-1 \,.
\end{equation}

\noindent We keep the notation $E_m;\;m=1,\ldots,N$, for the 
eigenvalues of $h$, with this convention they will be ordered from smallest 
to greatest.

From now on, we will pay particular attention to chains with $N/2$ an even 
number. 
As we will see below, we can state that, for

\begin{equation}
 \label{eses2}
N\,=\,2\,n\,=\, \alpha \,2^k \quad;\quad k \in  \mathbb{N} \,,
\end{equation}

\noindent  and $\alpha$ takes one of the values

\begin{equation}
 \label{efpa}
        \alpha\,=\,\left\{\begin{array}{l}
                2 \\
                3^p \times 5^q \times 17^r \times 257^s \times 65537^t 
                \quad;\quad
        p,\,q,\,r,\,s,\,t\,=\,0,1 \,, \end{array}\right.
\end{equation}

\noindent  The list of elements  of
$\alpha$ values smaller than $1000$ reads as follows

\begin{equation}
 \label{eamm}
\alpha\,=\,2,\,3,\,5,\,15,\,17,\,51,\,85,\,255,\,257,\,771\,.
\end{equation}

We want to remark here that. for these values of $\alpha$, there is an 
algebraic expression for the cosine functions in
Eq. \eqref{eavptnp} (see Appendix \ref{apcos}). This  allow us to 
write down algebraic expressions for all the eigenvalues and eigenvectors 
coefficients on which the TP depends. 

The following section deals with some particular cases of $\alpha$. We will pay 
close attention to the smallest values of this list, $\alpha=2,3$ and $5$, as a 
showcase of the methods employed to analyze chains of moderate length ($N\simeq 
20, 30)$. These methods rely on the Dirichlet's Approximation  Theorem for 
irrational numbers and the algebraic representation of particular values of 
trigonometric functions.

\section{Eigenvalues, Eigenvectors and Transmission Probabilities} 
\label{sec:examples}

In the previous section, we found analytical and simple expressions for the 
matrices corresponding to centrosymmetric Heisenberg spin chains such that 
$J_{2i-1}=1$ and $J_{2i}=J_2$ for $i =1,\cdots, N/2  \,;\quad N \in  
2\mathbb{N}$. We also showed that these
spin chains, with $N=\alpha \,2^k$, $k \in  \mathbb{N}$ and $\alpha$
taking values as in Eq. \eqref{efpa}, are exactly solvable. It is worth noting 
that, by exactly-solvable, we mean that both eigenvalues and the eigenvectors 
have algebraic expressions. The consequences of having algebraic expressions for 
the eigenvalues will be clear once we introduce Dirichlet's Approximation 
Theorem. We will dedicate this section to showing the explicit 
form of eigenvalues and eigenvectors of some of these spin chain
Hamiltonians, and we will use them to analyze the transmission probability by 
constructing analytical expressions for $P(t)$ defined in Eq. 
\eqref{eq:population}. In the following we will use transferred population 
and transmission probabilty as equivalent terms. We can not take a look at all 
the possible spin chains,  
and because of that, we will pay particular attention to small values of 
$\alpha$ ($\alpha=2,3$ and $5$) and some values of $k$.

\subsection{Chains with $N= 2 \times 2^k$ ($\alpha=2$, $n=2^k$)}

We will start with this relevant case, because it is known that it presents PGT
for $J_2=1$ \cite{Banchi2017}.  If we know the eigenvalues of 
$\mathfrak{h}^{(N/2)}$, then, using Eq. \eqref{ean2}, we only need to 
calculate $N/2=n$ new eigenvalues given by Eq. \eqref{eavptnpam},

\begin{equation}
\label{eavamuc}
\lambda_\pm^{(m)}\,=\, \left\{
\pm \,2\,\sqrt{1+J_2^2 \,+\, 2\,J_2\,\cos\left(\frac{(2\,m+1)\,\pi}{2^k}\right)}
        \quad;\quad m=0,\ldots,2^{k-1}-1 \right\}
\,,
\end{equation}

\noindent or, equivalently, using the algebraic expression of the cosine
functions (see Appendix \ref{apcos}),

\begin{equation}
\label{eavam}
\left\{\lambda_\pm\right \}_1^{2^{k-1}}\,=\,\left\{
\pm 2\,
\sqrt{1+J_2^2 \,+\, J_2  \overbrace{\sqrt{2\,\pm\,\sqrt{2\pm\ldots 
\pm \sqrt{2}}}}
^{k-2 \mbox{ nested square roots }}} \right\}\,,
\end{equation}

\noindent with $N=4$ as "initial condition", where all the eigenvalues are
calculated explicitly.
For this case, the Heisenberg Hamiltonian  is given by

\begin{equation}
 \label{ehamiltonian4}
 H = - ( \vec{\sigma}_1 \cdot \vec{\sigma}_2 + J_2 \; \vec{\sigma_2} \cdot
\vec{\sigma_3} +  \; \vec{\sigma_3} \cdot \vec{\sigma_4} )\,.
\end{equation}

\noindent and we can write our Hamiltonian in the one excitation 
subspace as follow:

\begin{equation}
\label{emh}
h^{(4)}\,=\,
\begin{pmatrix}
- J_2 & -2 & 0 & 0  \\
-2 & J_2 & -2 J_2  & 0  \\
0 & -2 J_2 & J_2 &  -2   \\
 0 & 0 & -2 & -J_2  
\end{pmatrix} 
\;\Rightarrow\;\mathfrak{h}^{(4)}\,=\,
\begin{pmatrix}
- 2\,J_2 & -2 & 0 & 0  \\
-2 & 0 & -2 J_2  & 0  \\
0 & -2 J_2 & 0 &  -2   \\
 0 & 0 & -2 & -2\,J_2
\end{pmatrix} 
\,.
\end{equation}

The eigenvalues of $\mathfrak{h}^{(4)}$ are

\begin{equation}
 \label{ehrN4}
\lambda_-^{(0)}=-2-2\,j2 \quad ;\quad \lambda_-^{(1)}=-2\,\sqrt{1+J_2^2}
 \quad ;\quad  
\lambda_-^{(2)}=-2+2\,j2  \quad ;\quad \lambda_+^{(1)}=2\,\sqrt{1+J_2^2} \,.
\end{equation}

\noindent Eigenvalues and eigenvectors of $h^{(4)}$, with the eigenvalues
ordered from smallest one to largest one are:


\begin{subequations}
\label{eaest4}
\begin{equation}
\label{eaest41}
E_1 \,=\,- 2 -J_2 \quad;\quad |E_1\rangle\,=\,\frac{1}{2} (1,1,1,1)\,,
\end{equation}
\begin{equation}
\label{eaest42}
E_2 \,=\,J_2-2 \sqrt{1+J_2^2}\quad;\quad |E_2\rangle\,=\,C_2\,
\left(-1,J_2- \sqrt{1+J_2^2},-J_2+ \sqrt{1+J_2^2},1 \right)
\end{equation}
\begin{equation}
\label{eaest43}
E_3 \,=\,2 -J_2 \quad;\quad |E_3\rangle\,=\,\frac{1}{2} (1,-1,-1,1)\,,
\end{equation}
\begin{equation}
\label{eaest44}
E_4 \,=\,J_2+2 \sqrt{1+J_2^2}\quad;\quad |E_4\rangle\,=\,C_4\,
\left(-1,J_2+ \sqrt{1+J_2^2},-(J_2+ \sqrt{1+J_2^2}),1 \right)
\end{equation}


\noindent where

\begin{equation}
\label{eaestc24}
C_{{2 \atop 4}} \,=\,\frac{1}{2 \sqrt{1+J_2^2\mp J_2 \sqrt{1+J_2^2}}}
\end{equation}
\end{subequations}

We can express the site basis as a linear combination of the eigenstates 
of the system. Then $|1\rangle$ and  $|4\rangle$ should read as:

\begin{eqnarray}
\label{e1ab}
	|1\rangle\,&=&\,\frac{1}{2} |E_1\rangle- C_2 |E_2\rangle+\frac{1}{2} 
	|E_3\rangle
-C_4 |E_4\rangle \\ \nonumber
	|4 \rangle\,&=&\,\frac{1}{2} |E_1\rangle+C_2 |E_2\rangle +\frac{1}{2} 
	|E_3\rangle 
+C_4 |E_4\rangle \,.
\end{eqnarray}

Finally, using these expressions and, with the help
of Equations \eqref{eaest4}, we can calculate 
$P(t)=|\langle1|U(t)|4\rangle|^2$ 

\begin{equation}
\label{ep4}
\begin{array}{ll}
P(J_2;t)&=\,\frac{1}{8} \left\{\cos \left[4\,t\right] +
\frac{2+3 J_2^2+\cos \left[4 \sqrt{1+J_2^2}\,t\right]}{1+J_2^2}
\right. \\
\mbox{} &\mbox{}\\
\mbox{}&\left. -\frac{
\cos \left[2 (1+J_2-\sqrt{1+J_2^2})\,t\right] +
\cos \left[2 (1-J_2+\sqrt{1+J_2^2})\,t\right]}{
1+J_2^2-J_2\,\sqrt{1+J_2^2}}
\right. \\
\mbox{} &\mbox{}\\
\mbox{}&\left. -\frac{\cos \left[2 (1+J_2+\sqrt{1+J_2^2})\,t\right] +
\cos \left[2 (1-J_2-\sqrt{1+J_2^2})\,t\right]}{
1+J_2^2+J_2\,\sqrt{1+J_2^2}}
\right\}\,.
\end{array}
\end{equation}

The PGT condition is satisfied if always exists a time $t_\varepsilon$, such 
that the value of the two cosine functions preceded by a 
plus sign are as close to the unity (the value of the four cosine functions 
preceded by a minus sign is as close to minus one) as required to obtain 
$P(t_\varepsilon)>1-\varepsilon$.  Banchi and collaborators \cite{Banchi2017} 
proved
that the case $J_2=1$ has PGT.  However, they did not give any estimation for  
$t_\varepsilon$ and, for this reason, we include in our study the homogeneous 
chain.
 
In particular, to obtain PGT it is necessary that

\begin{equation}
\label{epgtN4}
\cos(4\,t_\varepsilon)\,\simeq \,1 \;\Rightarrow\;t_\varepsilon\,\simeq\,
\frac{m}{2} \pi \;\Rightarrow\; 2\,\sqrt{1+J_2^2}\,m\,\simeq \,l \quad;\quad
\left( 1\pm J_2\pm \sqrt{1+J_2^2} \right)\,m\simeq 2 l_{\pm \pm}+1 \,.
\end{equation}

\noindent where $l$ y $l_{\pm \pm}$, and $m$ are natural numbers.

Dirichlet's Approximation Theorem ensures the fulfillment of these conditions. 
The theorem says that

\begin{teo}{ (Dirichlet's Approximation Theorem)} If $x_1,\dots,x_m\in \mathbb 
R$ 
and $M\geqslant 1$ is an integer, then there exists
an integer $q$ with $1 \leqslant q \leqslant  M^m$ and integers
 $p_1,\dots,p_m$ such that
$|q\,x_i-p_i|<1/M\quad;\quad i=1,\ldots,m$.
\end{teo}

The significance of this Theorem for our work is paramount.  It assures us that 
some natural numbers fulfill the conditions in Eq.~\eqref{epgtN4}  and that the 
PGT time is $t_{\varepsilon}=q \pi$.  Moreover, it is clear that the 
approximation of the irrational numbers is better for larger values of $M$. In 
other words, the larger the numbers $p_i$ and $q$ are considered, the better 
approximation results for the irrational numbers and the transmission 
probability.

The results about the PGT time included later in this work depend on the finding 
of successive approximations for the irrational numbers involved in the 
analytical expression of the transmission probabilities. 

As shown in Appendix~\ref{apcos} (by Niven's Theorem\cite{niven}) all the 
eigenvalues 
on which the transmission probabilities depend are functions of irrational 
numbers, see Eqs. 
\eqref{eavptnp} and \eqref{eavptnpam}. This dependency and the Dirichlet 
theorem give a systematic way to study chains with different lengths and values 
of the parameter $J_2$.  

In what follows, we will continue dealing with the spin chain with only four 
spins, clarifying the use of the Theorem.

Note that, accordingly with \eqref{epgtN4} for $J_2$ equal to a 
natural number, the arguments of the cosine functions contain only a single 
irrational number $\sqrt{1+J_2^2} = \sqrt{2}, \sqrt{5}, \sqrt{26}, \ldots$ for 
$J_2=1,2,5, \ldots$, respectively. In this simple case, the theorem assures 
that 
there are natural numbers $p$ and $q$ such that $ |q \sqrt{1+J_2^2}- p|<1/M $.

For $J_2=1$, the condition that the 
second cosine function has a value close to the unity implies that 
\[
 4 \sqrt{2} t_{\varepsilon} = r \pi , 
\]
with $r$ an even number then, if $t_{\varepsilon} \simeq q \pi/2$ (that is, 
the condition that makes that the first cosine function has a value close to 
the unity) the second cosine function argument can be rewritten as
\[
 4 \sqrt{2} \, q \frac{\pi}{2} \simeq 2p \pi ,
\]
which leads to
\[
 \sqrt{2} q \simeq p .
\]
The Theorem guarantees that this last equation is effectively fulfilled. 
It is rather direct to show that the conditions imposed by the other cosine 
functions that can be found in  \eqref{epgtN4} are also fulfilled once a pair 
$(p,q)$ is found. 
So, it is clear that for a given pair of values $(p,q)$ such that $|q \sqrt{2}- 
p|<1/M $ the time $t_{\varepsilon}= q \pi/2$ is a natural candidate to 
observe PGT. The actual value of $\varepsilon$ is calculated using that 
$\varepsilon=1-P(t_{\varepsilon})$. 

There is a small number of algorithms to actually obtain the $(p,q)$ pairs. 
Fortunately, the best known is the one that allows to find rational 
approximations for $\sqrt{2}$. The  best rational approximations for 
$\sqrt{2}$ are given
by the Newton-Rapshon succession,

\begin{equation}
\label{esr2}
a_0=1\quad;\quad a_{j+1}=\frac{a_j}{2}+\frac{1}{a_j} \quad ;\quad
a_j=\frac{p_j}{q_j} \;\Rightarrow\;
p_j=Num(a_j)\;;\;q_j=Den(a_j) \,,
\end{equation}

\noindent the first seven values of $p_j$, $q_j$  are shown in Table 
\ref{new-tpgtN8}

\begin{table}[!h]
\begin{center}
\begin{tabular}{c|c|c|c}
 $j$& $p_j$&  $q_j$ &$|q_j- \sqrt{2} p_j| $\\ \hline
 0&  1&  $1$ & 0.41421\\
  1&  3& $2$ & 0.17157\\
 2& 17&   $12$ & 0.02944\\
 3& 577&   $408$ &$8.7\times 10^{-4}$\\
 4& 665857&   $470832$ & $7.5 \times 10^{-7}$ \\
 5& 886731088897&   $627013566048$ \\
 6&$\quad$1572584048032918633353217& \quad $1111984844349868137938112$ 
\end{tabular}
\end{center}
\caption{The successive approximations to $\sqrt{2}$ as given by the 
Newton-Raphson succession. The index $j$ corresponds to the number of times 
that the recursive relationship in Eq.~\eqref{esr2} is employed. The quality 
of the approximation for the irrational number is remarkable as shown by the 
figures in the fourth column. 
\label{new-tpgtN8}}
\end{table}

The inclusion of this table here obeys a few reasons. The first one is related 
to the fact that the square root of two is an irrational number that appears in 
the expression of the transmission probabilities of chains with different 
lengths. The second, and more important, is that it manifests how 
$t_{\varepsilon}$ could grow quite fast. Of course, until the calculation of 
actual values of the transmission probabilities takes place, it is impossible 
to ascertain if these  times are needed to observe PGT for small enough values 
of $\varepsilon$. 

For other irrational numbers of the form $\sqrt{i}$, where $i$ is a natural 
number larger than or equal to two we use the continued fraction method to find 
the 
necessary $(p,q)$ pairs such that $p/q \simeq \sqrt{i}$.  
In particular, in the Tables \ref{tN4J21},  
\ref{tN4J22}, \ref{tN4J25}, and \ref{tN4J210} we show values of $p,q,P(J_2;t)$ 
for 
$J_2=1,\,2,\,5,$ and 10, see Appendix~\ref{ap:tables}. Note that the values of 
$p/q$ are approximations for 
$\sqrt{1+J_2^2}$. 

\begin{figure}[hbt]
\includegraphics[width=0.65\linewidth]{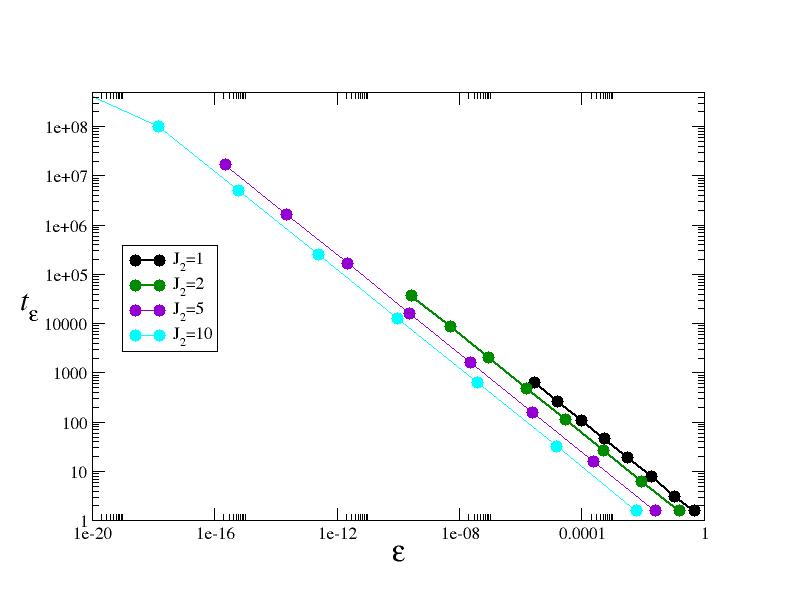}
\caption{The figure shows the behavior of the PGT time $t_\varepsilon$ vs. 
$\varepsilon$ for $N=4$ and different values of $J_2$ using a $\log - \log$ 
scale. The data points corresponding to  $J_2=1,2,5$ and $10$ are 
depicted using black, green, violet and cyan dots, respectively. The lines are 
included as a guide to the eye. Some of the data values are tabulated in  
Tables \ref{tN4J21} and \ref{tpgtN16}. 
\label{ftiempo} }
\end{figure}

\begin{figure}[hbt]
\includegraphics[width=0.65\linewidth]{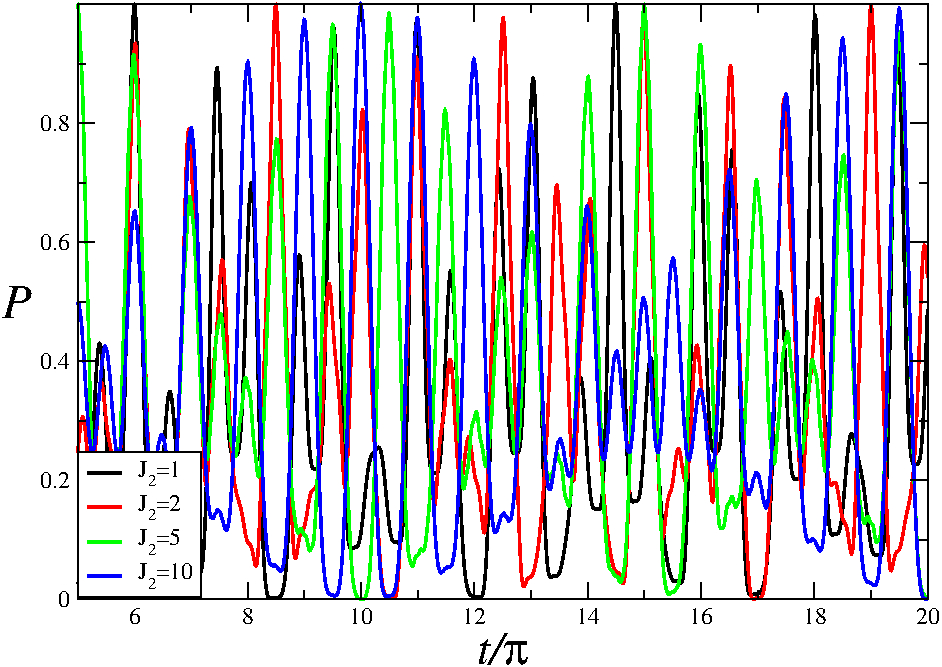}
\caption{The figure shows the behavior of $P(J_2;t)$ vs. $t/\pi$ for $N=4$, 
calculated using Eq. \eqref{ep4}. The  values of $P$ corresponding to 
$J_2=1,2,5$ and $10$ are shown as black, red, green and blue continuous lines, 
respectively. The time interval used in the figure was selected to include some 
of the  
values of $P(t_{J_2;\varepsilon})$ tabulated  in Tables \ref{tN4J21} and 
\ref{tN4J210}.}
\label{fcompN4}
\end{figure}

The results obtained for the PGT time $t_{\varepsilon}$ {\em vs} $\varepsilon$ 
for a chain of four spins and different values of $J_2$ are shown in 
Fig.~\ref{ftiempo}, while the  
transmission 
probabilities $P(J_2;t)$  for these values of $J_2$ 
are  shown in  Fig. \ref{fcompN4}.

As can be appreciated, the log-log scale used to show the data allow to observe 
a power law behavior of $t_\varepsilon$ vs. $\varepsilon$. Because the chain is 
quite short the excellent quality of the QST allow to reach 
very small values of $\varepsilon$. More interestingly, the power law exponent 
is clearly the same independently of $J_2$ for the values considered. 
Fitting each data set in Fig.~\ref{ftiempo} with the function 
\[
 t_\varepsilon = \frac{c}{(\varepsilon)^{f(N)}} ,
\]
results in $f(4)=1/2$. The value of the constant $c$ changes from data set to 
data set, but in the rest of the paper we will be interested only on the 
 the exponent, which is the relevant quantity.

On the 
other hand, these results show the ductility of the continued fraction method 
to find the necessary approximations for the irrational numbers involved in the 
calculation. Later on, we will show that the power law behavior can be observed 
for all the chains lengths that we investigated as long as PGT can be achieved.

As shown in Figure~\ref{ftiempo}, the PGT times necessary to achieve 
transmission probabilities very close to the unity are pretty short. For 
instance, for epsilon on the order of 0.0001, the PGT times approximately vary 
between 10 and 100. Shorter PGT times correspond to larger $J_2$ values. In 
figure 2, we plot the transmission probabilities as functions of time in an 
interval that allow us to appreciate the closeness to the unity achieved by each 
curve. However, Figure~\ref{fcompN4} shows how difficult it is to numerically 
estimate if PGT is achievable in a given chain by looking at the time behavior 
of the transmission probabilities.

\hspace{3mm}

\noindent {\em Chain with EC values that prevent the appearance of PGT} \\

It is interesting to note that our method allows us to ascertain that we are 
dealing with the PGT regime and calculate explicitly the characteristic time 
$t_{\varepsilon}$, for chains with different values of $J_2$. Nevertheless, 
following a similar procedure allow us to show that there are very particular 
values of $J_2$ where this kind of QST´ is forbidden.

By its definition, we know that if $P(J_2;t)$ presents PGT, then it is not a 
periodic function of $t$. Then, periodic QST is perfect, or otherwise, the 
transfer is poor. Kay \cite{Kay2019} showed that there is no PT in Heisenberg 
chains for $N>2$, so periodic chains must have a bad transmission.

Periodicity in $P(J_2;t)$ is obtained for  $J_2=b/a$, where $a<b<c$ is a 
Pythagorean triple, $a,\,b,\,c\;\in \mathbb{N}$ and $a^2+b^2=c^2$. In this case, 
we obtain $\sqrt{1+J_2^2}=c/a \;\in\,\mathbb{Q}$, and  $P(b/a;t)$ is a 
periodic function of $t$. For the smallest Pythagorean triple, $a=3,\,b=4,\,
c=5\;\Rightarrow\; J_2=4/3$ and $\sqrt{1+J_2^2}=5/3$ in Eq. \eqref{ep4} gives
\begin{equation}
\label{ep43}
P(4/3;t)\,=\,\frac{33}{100}+\frac{1}{10} \cos (4 \,  t)+\frac{9}{200}
    \cos \left(\frac{20 \,  t}{3}\right)
-\frac{9}{40} \left( \cos \left(\frac{4 \,  t}{3}\right)+\cos
    \left(\frac{8 \,  t}{3}\right)\right)-\frac{1}{40} \cos (8 \, t)
\end{equation}

\begin{figure}[hbt]
\includegraphics[width=0.65\linewidth]{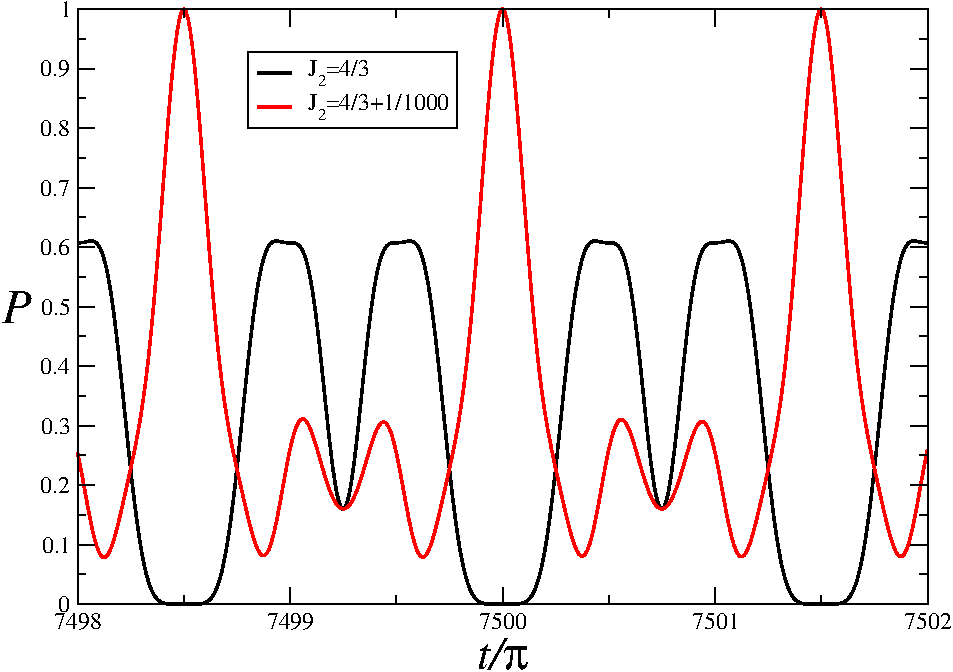}
\caption{$P(J_2;t)$ vs. $t/\pi$  calculated using Eq. \eqref{ep4}, for a 
spin chain with $N=4$. The transmission probabilities shown correspond to two 
particular values of the coupling $J_2$, the one linked to the
lowest Pythagorean triple, $J_2=4/3$ and  $J_2=4/3+1/1000$. The probabilities 
for these values of $J_2$ are shown using  black and red continuous lines, 
respectively. The time interval selected was chosen to show the behavior of 
both probabilities near $t_j=1500\times q_j\times \pi$ with $q_j=5$, where 
$P(4/3+1/1000,t)$ must be close to the unity, see Table 
\ref{tN4J243}.}\label{fN4J243}
\end{figure}

We obtain an upper bound for $P(4/3;t)$ replacing the cosines preceded by a 
plus sign by 1, and the cosines preceded by a minus sign by $-1$,  $P(4/3;t)
\leqslant 19/20=0.95$, and then, there is no PGT in this case. The true 
maximum is lower than 19/20, because for this bound we replace $ \cos (4 \,  t)$
by $1$ and $ \cos (8 \,  t)$ by $-1$. Actually, in this case the period of $P$
is $3\,\pi/2$, and the maxima can be
obtained analytically, and the first maximum is located at  
\begin{equation}
t_{max}=3\,\cos^{-1}\left(\sqrt{2+\sqrt{3/2}}/2\right)\simeq 0.4353 
\;\Rightarrow\;
P(4/3;t_{max})=625/1024=0.6103515625 ,
\end{equation}
see Fig. \ref{fN4J243}. 

If  $a,b,c$ are coprimes it is called a primitive Pythagorean triple
(non primitive triples have the expression
$(p\,a,\,p\,b,\,p\,c)$ where $p\,\in \,\mathbb{N}$). In this case 
$P(J_2=b/a,t)$ is a periodic function of $t$, and
we can prove that it does no present PGT as follows.
We will see that there is no $t^*$ such that the cosine functions  preceded 
with a plus sign approximate to 1, and  those preceded by a minus sign 
approximate to -1. Since there is always the term with $\cos(4 t)$, it
must be $t^*/\pi=q\,\delta(\varepsilon)/2$ where $q \,\in\,\mathbb{N }$ 
and $\delta(\varepsilon)$ is as close to 1 as we need.
  The arguments of the four cosine functions preceded by a
minus in \eqref{ep4} evaluated in $t^*$ are a factor $\pi$ multiplied by

\begin{equation}
\label{eacm1}
(a+b+c)\frac{q\,\delta(\varepsilon)}{a} \quad;\quad
|(a+b-c)|\frac{q\,\delta(\varepsilon)}{a} \quad;\quad
|(a-b+c)|\frac{q\,\delta(\varepsilon)}{a} \quad;\quad
|(a-b-c)|\frac{q\,\delta(\varepsilon)}{a} \,.
\end{equation}

\noindent Note that we include modules in order to have natural numbers, but 
the signals are not relevant. All these quantities in Eq. \eqref{eacm1} must
be close to  odd numbers in order to have PGT, that is

\begin{equation}
\label{eacm2}
\begin{split}
(a+b+c)\,q\,=\,(2\,r_{++}+1) \,a \quad &;\quad
|(a+b-c)|\,q\,\,=\,(2\,r_{+-}+1) \,a \quad;\\
|(a-b+c)|\,q\,\,=\,(2\,r_{-+}+1) \,a \quad &;\quad
|(a-b-c)|\,q\,\,=\,(2\,r_{--}+1) \,a \quad;\quad
\end{split}
\end{equation}
 
\noindent where $r_{\pm\pm}$ are natural numbers. 
It is known that or $a$ or $b$ is even and the other one and $c$ are odd 
numbers. Also we know that  $a\pm b\pm c$ are even numbers, two of them
$s_{1,2}$, are  divisible by 4, the other two, $s_{3,4}$ are not. 
Then , it is  clear  that Eqs. \eqref{eacm2} can not be fulfilled if $a$ is 
odd.  If $a$ is even,  then $a$ is divisible by 4, $a=4\,\tilde{a}$. 
For  $s_3$ (or $s_4$) we have
\begin{equation}
2\,(2\,l_3+1)\,q=4 \,(2\,r_3+1)\,\tilde{a} \;\Rightarrow\; (2\,l_3+1)\,q=
2 \,(2\,r_3+1)\,\tilde{a}  \;\Rightarrow\; q=2\,\tilde{q}. 
\end{equation}

\noindent Now we use this
result for $s_1$ (or $s_2$), $4 (s_1/4) \,2\,\tilde{q}= 
2 \,(2\,r_1+1)\,\tilde{a}$, condition that can not be fulfilled if $\tilde{a}$
is odd, and so on. Therefore, we prove that PGT is absent in a numerable set 
of values of $J_2=b/a$  where $a\,<\,b$ are the two smallest numbers of
 a primitive Pythagorean triple, and $P(J_2;t) $ is a periodic function of $t$. 
We will not develop the case $J_2=a/b<1$, which can be treated in the same way

Then, we have a numerable set of values where there is no PGT. It is 
interesting to study a neighborhood of one of this isolated points. In Table
\ref{tN4J243} we show values for $J_2=4/3+1/1000=4003/3000$ and, in Fig. 
\ref{fN4J243}, we compare $P(J_2;t)$ for $J_2=4/3$ and $4/3+1/1000$.

\begin{table}[h!]
\begin{center}
\begin{tabular}{c|c|c|c}
 $j$& $p_j$&  $q_j$ & $P(J_2=4/3+1/1000;t_j=1500\,q_j\,\pi)$  \\ \hline
 1& 5002  &  $1$ & 0.57264\\
 2& 10005 & 2 &0.8742 \\
  3& 25012 & 5 & 0.9999912 \\
 5&  24716860&  4941& 0.999999955\\
 6& 27803341 & 5558 & 0.999999982 \\
 8& 80323542  & 16057 & 0.9999999973 \\
9& 132843743 & 26556 & 0.99999999917 \\
11 & 346011028 & 69169 & 0.999999999964
\end{tabular}
\end{center}
\caption{Values of  $p_j,\;q_j$ and $P(4/3+1/1000;t_j)$ for a spin chains with 
$N=4$. The index $j$ gives the order of the continued fraction approximation for
$\sqrt{25024009}$ as $p_j/q_j$. The  values corresponding  to orders
$j=4, 7$ and $10$ do not meet the parity condition, and are not tabulated since 
they approximate zeros
of $P(4/3+1/1000;t)$ .\label{tN4J243}}
\end{table}

For chains of length $N=2\,\cdot\,2^k\,;\;k>1$ it is straightforward to see 
that, $\forall\;J_2>0, \;P(J_2;t)$ is not a periodic function of $t$.

 Looking at Fig. \ref{fcompN4} we can see how, when the coupling $J_2$ 
varies, the behavior of the transmission probability presents interesting 
changes. In particular, for $J_2$ large and $t\simeq n\pi/2$ with $n \in  
\mathbb{N}$ we see that the transmission reaches values very close to the 
unity. This strongly 
suggest that in the limit of strong coupling, we could explicitly show the 
appearance of PGT for particular Heisenberg spin chains. We are going to 
pay 
attention 
to this limit for any chain size in the next section of the present work.

\subsection{Chains with $N= 3\times2^k$ ($\alpha=3$)}

In this case, the initial condition for the recurrence equation \eqref{ean2}
is $N=6$ corresponding to $k=1$, that will be calculated explicitly in the
present section.
For $k>1$, we need to find the $3\times 2^{k-1}$ eigenvalues of
$A_-^{(N/2)}$. In order to simplify this task, we separate the  eigenvalues 
given  by Eq. \eqref{eavptnpam}
 into two sets, depending on whether $(2\, l + 1)$ is a multiple 
of 3 or not,

\begin{equation}
\label{eavamuc3b1}
\left\{\lambda_\pm\right\}_1^{ 2^{k-1}}\,=\left\{ \pm\,2\,
        \sqrt{1+J_2^2 \,+\, 2\,J_2\cos\left(\frac{(2\,m+1)\,\pi}{2^{k-1}}
        \right)}\quad;\quad m=0,\ldots, 2^{k-2}-1\right\}
\,,
\end{equation}

\noindent  which are identical to the eigenvalues of $A_-^{(2^{k-1})}$, and

\begin{equation}
\label{eavamuc3b2}
\left\{\lambda_\pm\right \}_{2^{k-1}+1}^{3\times 2^{k-1}}\,=\,\left\{
\pm\,2\,
\sqrt{1+J_2^2 \,+\, 2\,J_2\cos\left(a_3(m)\,\frac{\pi}{3 \cdot2^{k-1}}\right)}
\quad;\quad m=0,\ldots, 2^{k-1}-1 \right\}\,,
\end{equation}

\noindent where $a_3(m)=((6\, m+1)-(-1)^m)/2+1$ represents the odd numbers 
that are not multiple of $3$.

Using the algebraic expression of the cosine functions we obtain,

\begin{eqnarray}
\label{eavam3}
	&&\left\{\lambda_\pm\right \}_1^{ 2^{k-1}}\,=\,\left\{ \pm\,2\,
\sqrt{1+J_2^2 \,+\, J_2  \overbrace{\sqrt{2\,\pm\,\sqrt{2\pm\ldots 
\pm \sqrt{2}}}}
	^{k-2 \mbox{ nested square roots}}} \right\} \;;\\ \nonumber
&&\left\{\lambda_\pm\right \}_{2^{k-1}+1}^{3\times 2^{k-1}}\,=\,\left\{\pm\,2\,
\sqrt{1+J_2^2 \,+\, J_2  \overbrace{\sqrt{2\,\pm\,\sqrt{2\pm\ldots\pm\sqrt{3}}}}
^{k-1 \mbox{ nested square roots}}}\right\}
\,.
\end{eqnarray}

Now we pay attention to the simplest case, $k=1\;\Rightarrow\;N=3\times 2^1=6$
and we write the corresponding Heisenberg Hamiltonian

\begin{equation}
 \label{ehamiltonian6}
 H = - ( \vec{\sigma}_1 \cdot \vec{\sigma}_2 + J_2 \; \vec{\sigma_2} \cdot
\vec{\sigma_3} +  \; \vec{\sigma_3} \cdot \vec{\sigma_4}  + J_2 \;
\vec{\sigma_4} \cdot \vec{\sigma_5}
+ \vec{\sigma_5} \cdot \vec{\sigma_6}) \,,
\end{equation}

\begin{figure}[hbt]
\includegraphics[width=0.7\linewidth]{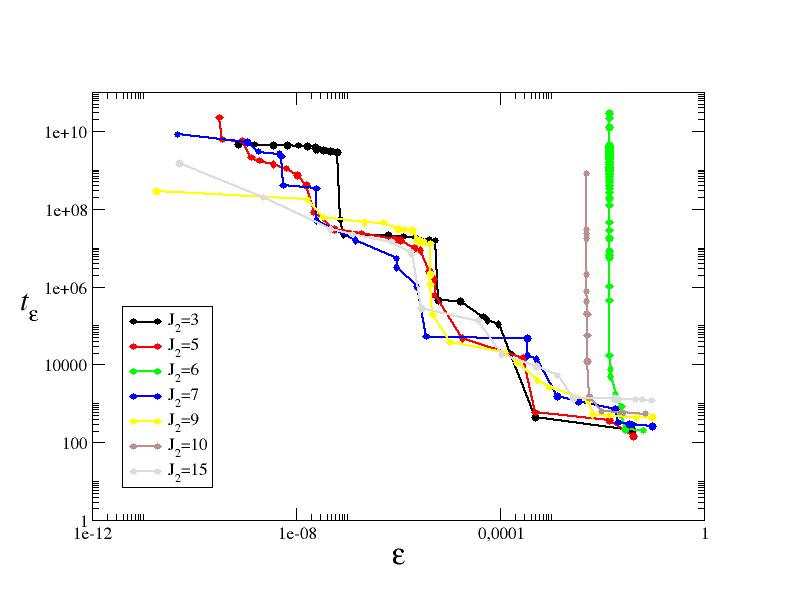}
\caption{ The PGT time $t_{\varepsilon}$ as a function of $\varepsilon$ for 
different values
	of $J_2$ for a chain with $N=6$ spins. The values of $t_{\varepsilon}$ are 
calculated using the continued fraction method previously described. These 
vales are shown as solid dots, while the lines are included as a guide. The 
data sets correspond to the two different possible behaviors described in the 
text. For $J_2=3,5,7,8$ and $15$ no matter how small $\varepsilon$ is, there is 
a corresponding time $ t_\varepsilon$ such that 
$1-P/(J_2;t_\varepsilon)<\varepsilon$. For $J_2=6$ and $10$ (green and brown 
dots, respectively) it does not matter how well the continued fraction method 
approximates the irrational numbers involved in the calculation of $P$. Even for 
larger and larger values of the time obtained from the Dirichlet's 
approximation method the respective values of $\varepsilon$ 
remain without decreasing their value. \label{fn6j2}}
\end{figure}

We can obtain algebraic expressions for all the eigenvalues and eigenvectors 
of this Hamiltonian. We also can compute, analytically, 
the localization of all the eigenstates in any site ($i$) of the spin chain. 

It is worth pointing out that the theorems that affirm that 
in chains such that $N=2^k$ PGT is  achievable, assume that the chains have 
homogeneous exchange 
couplings coefficients. Applying the algorithm already described to find 
rational approximations for irrational numbers, we  show numerical evidence 
that  in chains with $N=3\times 2^k$ PGT can be achieved by selecting 
appropriate 
values for $J_2$.

Figure~\ref{fn6j2} shows the values of the PGT time, $t_{\varepsilon}$, for a 
chain with six spins and  different values of the coupling $J_2$. As the 
data show, if $J_2$ is chosen as an odd integer bigger than one, the algorithm 
finds PGT times for arbitrarily small values of $\varepsilon$. On the other 
hand, for even values of $J_2$, it is not possible to reach arbitrarily close 
to $P=1$, no matter how long the arrival times considered are. See the vertical 
lines joining the successive approximations found by the algorithm for the 
irrational numbers involved in the formula of $P(t)$ for $N=6$. The 
green and brown vertical lines correspond to $J_2= 5$ and $10$. The algorithm 
used to find the required approximations for the irrational numbers is 
inefficient computationally, so obtaining PGT times larger than $10^{10}$ 
becomes quite expensive. This restricted the values of $\varepsilon$ that can 
be obtained numerically.

We fitted the data corresponding to odd values of $J_2$ with a power-law 
resulting in an exponent equal to the unity.  As we will show, this result is 
consistent with the 
exponents calculated for chains $N=2^k$ and sits between the exponents for $N=4$ 
and $N=8$.

\subsection{Chains with $N= 5\times2^k$ ($\alpha=5$)}

In this section we will calculate the cases corresponding to $N=5 \times 2^k$.
The initial condition, corresponding to $k=1$ is $N=10$, 
for $k>1$, the eigenvalues of $A_-^{(5\cdot 2^{k-1})}$, given by 
 Eq. \eqref{eavptnpam} are, as before, divided in two sets,

\begin{equation}
\label{eavamuc5b1}
\left\{\lambda_\pm\right\}_1^{ 2^{k-1}}\,=\,\left\{\pm\,2\,
\sqrt{1+J_2^2 \,+\, 2\,J_2\cos\left(\frac{(2\,m+1)\,\pi}{2^{k-1}}\right)}
\quad;\quad m=0,\ldots, 2^{k-2}-1 \right\} \,,
\end{equation}

\noindent which, as the case $\alpha=3$,  are identical to the eigenvalues of 
$A_-^{(2^{k-1})}$, and

\begin{equation}
\label{eavamuc5b2}
\left\{\lambda_\pm\right \}_{2^{k-1}+1}^{5\times 2^{k-1}}\,=\,\left\{\pm\,2\,
\sqrt{1+J_2^2 \,+\, 2\,J_2\cos\left(\frac{a_5(m)\,\pi}{5 \cdot2^{k-1}}\right)}
\quad;\quad m=1,\ldots, 2^{k}  \right\} \,,
\end{equation}

\noindent where 
\begin{equation*}
a_5(m)=(10\,m-9-(-1)^m+(1-i)\,
i^{-m}+(1+i)\,i^m)/4+1, \quad i=\sqrt{-1},
\end{equation*}
represents the odd numbers that are
not multiples of $5$. Replacing the cosine functions by their algebraic 
expression we 
obtain,

\begin{eqnarray}
\label{eavam5}
	&&\left\{\lambda_\pm\right \}_1^{ 2^{k-1}}\,=\,\left\{ \pm\,2\,
\sqrt{1+J_2^2 \,+\, J_2  \overbrace{\sqrt{2\,\pm\,\sqrt{2\pm\ldots\pm\sqrt{2}}}}
^{k-2 \mbox{ nested square roots}}} \right\} \;; \\ \nonumber
	&&\left\{\lambda_\pm\right \}_{2^{k-1}+1}^{3\times 2^{k-1}}\,=\,\left\{
\pm\,2\,
\sqrt{1+J_2^2 \,+\, J_2  \overbrace{\sqrt{2\,\pm\,\frac{\sqrt{2\pm\ldots
 \pm\sqrt{\frac{2+\sqrt{5}}{2}}}}{2}}}
^{k-1 \mbox{ nested square roots}}} \right\}
\,.
\end{eqnarray}

\noindent where $m=1,\ldots, 2^{k}$. It is possible to write the eigenvalues, 
eigenvectors 
and even the transmission probability but they are very long and complicated 
expressions  to transcribe them in a paper.

\begin{figure}[hbt]
\includegraphics[width=0.75\linewidth]{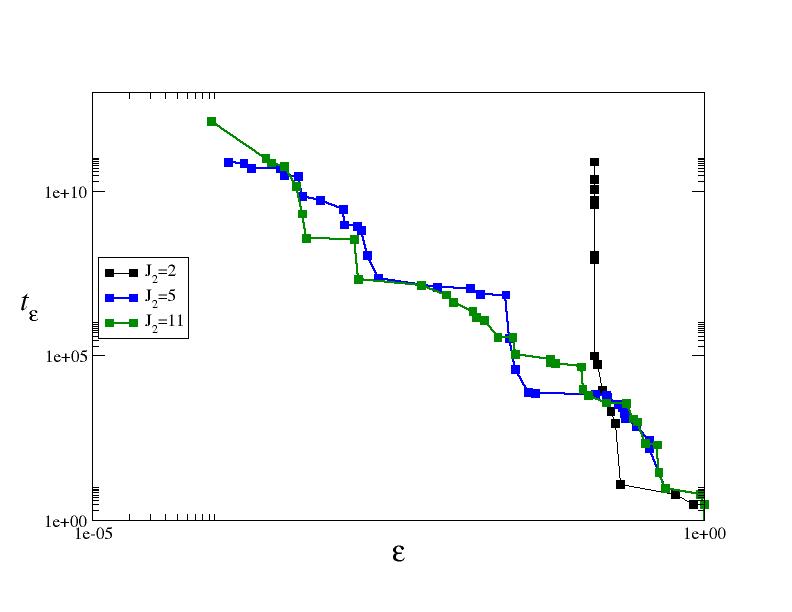}
\caption{The PGT time $t_{\varepsilon}$ as a function of $\varepsilon$ for 
different values
	of $J_2$ for a chain with $N=10$ spins. The values of $t_{\varepsilon}$ are 
calculated using the continued fraction method previously described. These 
vales are shown as square dots, while the lines are included as a guide. The 
data sets correspond to the two different possible behaviors described in the 
text. For $J_2=5$ and $11$ no matter how small $\varepsilon$ is there is 
a corresponding time $ t_\varepsilon$ such that 
$1-P/(J_2;t_\varepsilon)<\varepsilon$. For $J_2=2$ and $10$ it 
does not matter how well the continued fraction method 
approximates the irrational numbers involved in the calculation of $P$. Even 
for 
larger and larger values of the time obtained from the Dirichlet's 
approximation method the respective values of $\varepsilon$ 
remain without decreasing their value.}
\label{fcompN10}
\end{figure}

\subsection{Eigenvectors and localization}

So far, the results presented for small chains strongly depend on the ability 
to produce rational approximation for the irrational numbers that appear in 
expressions for the transmission probabilities involving the difference 
between eigenvalues. Nevertheless, we have paid little attention to the 
behavior of the eigenvectors and its components. 

For the examples analyzed the 
role of the eigenvectors is not as relevant as a good approximation for the 
eigenvalues differences, even in the case when the value of $J_2$ prevents the 
appearance of PGT. In Appendix~\ref{ap:loc-eigenvectors} we analyze some 
properties of the eigenvectors of short spin chains.

\section{Perfect, Good and Pretty Good transmission in the limit of strong 
coupling} \label{sec:jinfty}

Achieving an efficient transmission on Heisenberg spin chains is of fundamental 
importance for quantum technologies and related issues. Up to this point, we 
have studied a family of quantum spin Heisenberg chains and found analytical 
expressions for several quantities, in particular the transmission 
probabilities. It is reasonable to use these results to answer questions related 
to the efficient transmission of states. As shown in the previous section,  it 
is possible to find, in some spin chains, conditions that allow for QST with 
high fidelity. In Fig. \ref{fcompN4} we found some values of time for which the 
transmission probability between the first and last site seems to be very high. 
This situation occurs when the strength of the coupling $J_2$ is large enough.

We start analyzing the limit $J_2 \rightarrow \infty$ for the $N=4$
spin chain studied in Section \ref{sec:examples},

\begin{equation}
\label{elimp4}
P_\infty(t)\,=\,\lim_{J_2\rightarrow \infty} P(t)\,=\, \frac{1}{8}
\left( 3+\cos \left(4\,t\right)- 4 \cos \left(2\,t\right) \right)
\,=\,\sin^4(t)\, .
\end{equation}

\noindent It is easy to check that we obtain perfect transmission (PT) in this 
limit for the times shown below

\begin{equation}
\label{elimptp4}
P_\infty(t_{PT}=(m-1/2) \pi)\,=\,1 \quad;\quad P_\infty(t_{null}=m \pi)\,=\,0
\,,
\end{equation}

\noindent where $m \in \mathbb{N}$. This result becomes relevant when we
note that, in general, there is no 
PT for Heisenberg chains in the case of arbitrary finite interaction for 
$N>2$. Therefore it results important to investigate spin chains with
$N>4$ in the limit of strong coupling. In order to solve our
problem for any value of $N$ we employ first order perturbation
theory in $\varepsilon=1/J_2$,

\begin{equation}
\label{edh}
h=h_1+J_2\,h_0=J_2\,(h_0+\eta \,h_1)\,.
\end{equation}
 
\noindent where matrices $h_0$ and $h_1$ have no free parameters. In the 
following we call the matrices $h_0$ and $h_1$ the zeroth and first order 
Hamiltonian matrix, respectively.
It is important to note that, as the eigenstates are degenerated
at $\eta\rightarrow 0$, we need to use degenerated perturbation
theory.

\subsection{Properties of zeroth order Hamiltonian matrix}\label{ph0}

Taking the limit $J_2\to \infty$ in Eq. \eqref{emhg} we obtain, for chains
with length $N=2\,n$


\begin{equation}
\label{emh0}
h_0\,=\,
\begin{pmatrix}
-n+1&  0&  0 & 0&0&0& \ldots &  0 &  0&0 \\
0 &  -n+3 &  -2& 0&0&0& \ldots &  0 &  0 &0\\
0 &  -2&  -n+3 & 0 &  0 &  0 & \ldots &  0 &  0&0 \\
0 &  0 &  0 &  -n+3 & -2&  0& \ldots &  0 &  0&0 \\
0 &  0&0&-2&  -n+3 &  0 & \ldots &  0 &  0 &0\\
\vdots&\vdots&\vdots&\vdots&\vdots&\vdots&\ddots&\vdots&\vdots\\
0 &  0&0&0&  0 &  0 & \ldots &  -n+3 &  -2&0  \\
0 &  0&0&0&  0 &  0 & \ldots &  -2 &  -n+3&0  \\
0 &  0 & 0&0& 0& 0 &\ldots &  0& 0& -n+1
\end{pmatrix} \,,
\end{equation}


For this matrix holds:

\begin{enumerate}[i)]

\item It is bisymmetric.

\item This matrix has $n-1$ identical blocks of size $2\times 2$ 
with eigenvalues 

\begin{equation}
\label{eph0}
E_<^{(0)}\,=\,-n+1  \quad;\quad E_>^{(0)}\,=\,-n+5 \,.
\end{equation}

\noindent and the corresponding eigenvectors are $(1,1)/\sqrt{2}$ and
$(1,-1)/\sqrt{2}$, and 	two identical one dimension blocks with 
eigenvalues $E_<^{(0)}$.It follows that the 
degeneration of these eigenvalues are:

\begin{equation}
\label{edegh0}
deg_<=n+1 \quad;\quad deg_>=n-1 \,.
\end{equation}

\noindent We call $S_<$ and $S_>$ to the subspaces expanded by the 
eigenvectors of $E_<^{(0)}$ and  $E_>^{(0)}$ respectively.

\item \label{iavh0} We can find explicit expressions of the eigenvectors 
of $S_<$,

\begin{subequations}
\label{eavech0}
\begin{equation}
\label{eavech012}
\hat{u}_1\,=\,(1,0,0,0,0,0,\ldots,0,0) \quad;\quad
 \hat{u}_2\,=\,\frac{1}{\sqrt{2}}(0,1,1,0,0,0,\ldots,0,0) \quad;
\end{equation}
\begin{equation}
\label{eavech034}
\hat{u}_3\,=\,\frac{1}{\sqrt{2}}(0,0,0,1,1,0,\ldots,0,0) \quad;\quad
 \hat{u}_4\,=\,\frac{1}{\sqrt{2}}(0,0,0,0,0,1,1,\ldots,0) \quad;
\end{equation}
\begin{equation}
\label{eavech0n}
\hat{u}_n\,=\,\frac{1}{\sqrt{2}}(0,0,0,0,\ldots,1,1,0) \quad;\quad
 \hat{u}_{n+1}\,=\,(0,0,0,0,0,0,\ldots,0,1) \quad ,
\end{equation}
\end{subequations}

\noindent then the restriction of $h_0$ to $S_<$ is
\begin{equation}
\label{emah0r}
h_0|_{S_<}\,=\,(-n+1)\; \mathbb{I},
\end{equation}

\noindent where $\mathbb{I}$ is the identity matrix of
dimension $n+1$. It is important to note that $|1\rangle$ y $|N\rangle $ 
are eigenstates of $h_0$ in $S_<$. 

Finally, the eigenvectors in the subspace  $S_>$ are:


\begin{subequations}
\label{eavech0b}
\begin{equation}
\hat{v}_1\,=\,\frac{1}{\sqrt{2}}(0,1,-1,0,0,\ldots,0,0) \quad;\quad
 \hat{v}_2\,=\,\frac{1}{\sqrt{2}}(0,0,0,,1,-1,\ldots,0,0) \quad; \ldots
\end{equation}
\begin{equation}
\hat{v}_{n-2}\,=\,\frac{1}{\sqrt{2}}(0,0,\ldots,0,1,-1,0,0,0) \quad;\quad
 \hat{v}_{n-1}\,=\,\frac{1}{\sqrt{2}}(0,0,0,0,\ldots,0,1,-1,0) \quad,
\end{equation}
\end{subequations}


\end{enumerate}

\subsection{Properties of the first order Hamiltonian matrix}\label{ph1}

The expression for this matrix consists in  $n$ $2\times 2$ matrices
in the diagonal,

\begin{equation}
\label{emah1}
h_1\,=\,
\begin{pmatrix}
a_2&  0&  0 & 0&0&0& \ldots &  0 &  0 \\
0 &  a_2 &  0& 0&0&0& \ldots &  0 &  0 \\
0 &  0&  a_2 &  0 &  0 &  0 & \ldots &  0 &  0 \\
\vdots&\vdots&\vdots&\vdots&\vdots&\vdots&\vdots&\vdots&\vdots\\
0 &  0&0&0&  0 &  0 & \ldots &  a_2 &  0 \\
0 &  0 & 0&0& 0& 0 &\ldots &  0&  a_2
\end{pmatrix} \,, \quad \mbox{ where }
a_2\,=\,
\begin{pmatrix}
-n+2&-2 \\
-2 &-n+2
\end{pmatrix}
\end{equation}

For this matrix holds:

\begin{enumerate}[a)]
\item Is bisymmetric.

\item The restriction of $h_1$ to $S_<$, of dimension
$n+1\times n+1$, is

\begin{equation}
\label{emah1r}
 h_1|_{S_<}\,=\,
\begin{pmatrix}
-n+2&   -\sqrt{2} & 0&0&0& \ldots&0 &  0 &  0 & 0\\
 -\sqrt{2} &  -n+2 &  -1& 0&0& \ldots &0 & 0 &  0 &  0 \\
0 &   -1& -n+2 &  -1 &  0  & \ldots &  0 &  0& 0& 0 \\
\vdots&\vdots&\vdots&\vdots&\vdots&\vdots&\vdots&\vdots&\vdots\\
0 &  0&0 &  0 &  0 & \ldots &-1&  -n+2 &  -1&  0 \\
0 &  0&0&  0 &  0 & \ldots &0 &-1&  -n+2 &    -\sqrt{2} \\
0 &  0 & 0&0&  0 &\ldots& 0  & 0 &  -\sqrt{2} &  -n+2
\end{pmatrix} \,,
\end{equation}

\item The eigenvalues of $h_1|_{S_<}$ are

\begin{equation}
\label{eave2k}
E_{m+1}^{(1)}\,=\, -n+2+2\,\cos\left(m\,\frac{\pi}{n}\right) \quad ; \quad
m=0,\,1,\,\ldots,\,n \,.
\end{equation}

\noindent It is possible to obtain an algebraic expression for these
eigenvalues only for the same cases described for 
finite values of $J_2$ as showed in Eq. \eqref{efpa}

\end{enumerate}

Using the properties of $h_0$ and $h_1$ described in Sections
\ref{ph0} and \ref{ph1}, we obtain the self energies of $h|_{S_<}$
to first order in perturbation:

\begin{equation}
\label{eavo1}
E_i=- J_2 (n-1)-n+2+2\,\cos\left((i-1)\,\frac{\pi}{n}\right)\,+\,
	{\cal O}\left( \frac{1}{J_2} \right)  \quad ; \quad i=1,\,2,\,
	\ldots,\,n+1\,.
\end{equation}

At this moment we have all the ingredients in order to calculate 
the transmission probability in the strong coupling limit.

\subsection{The transferred population as a function of time in the strong 
coupling limit}

In this section we obtain a very simple formula for the the PT in strong 
coupling limit 

\begin{equation}
\label{epinf}
P_\infty(t)=\lim_{J_2\rightarrow \infty}\left|\langle1|
e^{-i\,h(J_2)\,t}|N\rangle\right|^2=\sum_{i,j=1}^{n+1}\, (-1)^{i+j} \,w_{i}^2
	\,w_{j}^2\,
e^{-i\,(E_i-E_j)\,t}  
\,,
\end{equation}

\noindent where $w_i^2$ are the square of the coefficients of the expansion 
for $|1\rangle$ 
and $|N\rangle$ in the eigenbasis of $h$ in the strong coupling limit,

\begin{equation}
\label{ecwi}
w_{i}^2\,=\,\left\{ \begin{array}{ll}
\frac{1}{N} & \mbox{ if }i=1,\,n+1 \\
 & \\
\frac{1}{n}  & \mbox{ if }i \ne 1,\,n+1
\end{array} \right.\,,
\end{equation}

From Eqs. \eqref{epinf} and \eqref{ecwi} we get that

\begin{equation}
\label{epinfr}
\begin{split}
P_\infty(t) &=\,\left|\sum_{i=1}^{n+1}\, (-1)^{i} \,w_{i}^2\,
e^{-i\,2\,\cos((i-1)\,\pi/n)\,t} \right|^2 \\
&=\,\left[\sum_{i=1}^{n+1}\, (-1)^{i} \,w_{i}^2\,\cos\left(
2\,\cos((i-1)\,\pi/n)\,t \right) \right]^2 \,+\,
\left[\sum_{i=1}^{n+1}\, (-1)^{i} \,w_{i}^2\,\sin\left(
2\,\cos((i-1)\,\pi/n)\,t \right) \right]^2  \\
&=\,\frac{1}{n^2}\,\left\{
\left[ \frac{(1+(-1)^n)}{2}\cos(2 \,t)+\sum_{i=1}^{n-1}\,(-1)^{i} \,\cos\left(
2\,\cos(i\,\pi/n)\,t \right) \right]^2 \,+ \right. \\
&\left. \hspace{1.4cm} \left[ \frac{-(1-(-1)^n)}{2}\sin(2 \,t)+
\sum_{i=1}^{n-1}\,(-1)^{i} \,
\sin\left(2\,\cos(i\,\pi/n)\,t \right) \right]^2
\right\}
\,.
\end{split}
\end{equation}

\noindent Now we split this equation for $P_\infty$ for odd and even
values of $n$,

\begin{itemize}
\item{$n$ odd}

\begin{equation}
\label{epinfimpar}
P_\infty(t) \,=\,
\frac{1}{n^2} \;\left(\sum_{i=0}^{n-1} \, (-1)^i \,
\sin\left[2\;\cos\left(i\,\frac{\pi}{n}\right)\;t\right]\right)^2 \,=\,
\frac{1}{n^2} \;\left(\sin(2\,t)\,+\,2 \sum_{i=1}^{(n-1)/2} \, (-1)^i \,
\sin\left[2\;\cos\left(i\,\frac{\pi}{n}\right)\;t\right]\right)^2 \,.
\end{equation}

\item{$n$ even}

\begin{equation}
\label{epinfpar}
P_\infty(t) \,=\,
\frac{1}{n^2} \;\left(\sum_{i=0}^{n-1} \, (-1)^i \,
\cos\left[2\;\cos\left(i\,\frac{\pi}{n}\right)\;t\right]\right)^2\,=\,
\frac{1}{n^2} \;\left((-1)^{n/2}+\cos(2\,t)+2\sum_{i=1}^{n/2-1} \, (-1)^i \,
\cos\left[2\;\cos\left(i\,\frac{\pi}{n}\right)\;t\right]\right)^2 \,.
\end{equation}

\end{itemize}

\subsection{Non existence of perfect transfer in the strong coupling 
limit for chains with more than four spins}

In this part of the work we will employ Eq. \eqref{epinfimpar} 
that holds when we take $k=1$ and $\alpha$ odd in Eq. \eqref{eses2}, which
implies $n=\alpha$ odd. There exist two conditions in order to obtain PT,
one of these is $\sin{(2\,t_{PT})}=1\Rightarrow t_{PT}=(m+1/4)\,\pi$. It is
sufficient to analyze the first term of the sum in Eq. \eqref{epinfimpar}
which should obey $\cos(\pi/\alpha) \,(2 \,m+1/2)\,
\pi\,=\,(2\,l+3/2)\,\pi\Rightarrow \cos(\pi/\alpha) \,=\,(4\,l+3)/(4\,m+1)$.
By Niven's Theorem \cite{niven} we know that $\cos(\pi/\alpha)$ is a
rational number only for $\alpha=1,2$ and $3$. Then, we have
proved that there is no PT for $n>3$ odd. For the case $n=3$ we can 
calculate explicitly $P_\infty$:

\begin{equation}
\label{elimp6}
P_\infty(t)\,=\, \frac{16}{9}
\sin^4\left(\frac{t}{2}\right)\,\sin^2(t)\,,
\end{equation}

\noindent This transmission probability has its maxima

\begin{equation}
\label{elimptp7}
P_\infty(t_{max}=2\,(m-2/3) \pi)\,=\,P_\infty(t_{max}=2\,(m-1/3) \pi)\,=\,
\frac{3}{4} 
\quad;\quad m \in \mathbb{N} \,,
\end{equation}

\noindent which is far away form the unity. The second condition, 
$\sin{(2\,t_{PT})}=-1$, can be demonstrated
in a similar way. 

Now we need to check what happen in the case of $n$ even, corresponding
to $k>1$. With this aim we have to use Eq. \eqref{epinfpar}, but we need to
study two different cases, $n/2$ even and $n/2$ odd. For the first one, must
hold $\cos(2\,t_{PT})=1\;\Rightarrow\;t_{PT}=m\,\pi$. As before
it is sufficient to analyze the first term of the sum in Eq. \eqref{epinfpar},
$2\;\cos\left(\frac{\pi}{n}\right)\;t_{PT}=(2\,l+1) \,\pi \;\Rightarrow\;
\cos\left(\frac{\pi}{n}\right)= (2\,l+1)/2\,m$ but from 
Niven's Theorem \cite{niven} we know that $\cos\left(\frac{\pi}{n}\right)$
is an irrational number for $n\geqslant 4$.

Then, we have proven that there is no PT for chains with $N>4$ in the
large coupling limit.

\subsection{Pretty Good Transmission in the Large Coupling Limit}

In this section we will show that Heisenberg Spin Chains
with $\alpha=2$ present PGT. For chains with $\alpha\geqslant 3$ it is
straightforward to show, calculating explicitly from Eq. \eqref{epinfimpar}, 
that there is no PGT for $k=1$. For the sake of clarity we write down 
the cases $\alpha=3$ and $\alpha=5$ with $k=1$:

\begin{equation}
\label{elimp35}
	P^{(N=6)}_\infty(t)\,=\,\frac{16}{9}
\sin^4\left(\frac{t}{2}\right)\,\sin^2(t) \quad;\quad
	P^{(N=10)}_\infty(t)\,=\,
\frac{1}{25}
\left( \sin(2\,t) -4\,\cos(\sqrt{5}\, t/2)  \,\sin(t/2)\right)^2 \,,
\end{equation}

\noindent where the maxima are far away from the unity. We will assume
that there is no PGT for $k>1$. We checked this assumption in several cases
using numerical implementations of the corresponding  analytical 
expressions.

For the case $\alpha=2$ and arbitrary $k$ the Eq. \eqref{epinfpar} takes
the form

\begin{equation}
\label{epinf2k}
\begin{split}
P_\infty(t) &=\, \frac{1}{2^{2 k}}
	\left\lbrace 1 + \cos\left(2\,t \right
	)+ 2 \,\left[\cos\left(\sqrt{2} \;t \right) +
	\,\cos\left(\sqrt{2-\sqrt{2}}\;t\right)+
	\,\cos\left(\sqrt{2 + \sqrt{2}}\;t \right)  +   \right.  \right. \\
	&\left.  \cos\left(\sqrt{2 + \sqrt{2+\sqrt{2}}}\;t \right)+
	\cos\left(\sqrt{2 + \sqrt{2-\sqrt{2}}}\;t \right)+
	\cos\left(\sqrt{2 - \sqrt{2-\sqrt{2}}}\;t \right)+
	\cos\left(\sqrt{2 - \sqrt{2+\sqrt{2}}}\;t \right)+ \ldots + 
	\right. \\
	&\cos\left( \overbrace{
\sqrt{2+\sqrt{2+\sqrt{2+\ldots +\sqrt{2}}}}}^{k-2 \mbox{ nested square roots}}
\,t\right) 
\,+\,\cdots\,+\,     \left.
        \cos\left( \overbrace{
\sqrt{2-\sqrt{2+\sqrt{2+\ldots +\sqrt{2}}}}}^{k-2 \mbox{ nested square roots}}
\,t\right) 
	\right] \\
	& -2  \left[\cos\left( \overbrace{
\sqrt{2+\sqrt{2+\sqrt{2+\ldots +\sqrt{2}}}}}^{k-1 \mbox{ nested square roots}}
\,t\right) 
 \,+\,\cdots\,+\, 
\left. 
	\cos\left( \overbrace{
\sqrt{2-\sqrt{2+\sqrt{2+\ldots +\sqrt{2}}}}}^{k-1 \mbox{ nested square roots}}
\,t\right) \right] \right\rbrace^2 \,,
\end{split}
\end{equation}

\noindent where the last square bracket has $2^{k-2}$ terms corresponding to 
the $2^{k-2}$ possible combinations of positive and negative signs. In order
to have PGT we need a value of $t_{\varepsilon}$ such that the arguments of 
the cosine functions present in the first bracket should be as close as
required to even integers multiplied by $\pi$, and the arguments of
the cosine functions contained in the second bracket should be as close as
required to odd integers multiplied by $\pi$.

As before, $t_{\varepsilon}=q\pi$ assures that the modulus of all cosine 
functions
are as close to he unity as we require. Dirichlet's approximation theorem has 
not information about
the parity of $q$ and $\{p_i\}$ but nevertheless in all the cases analyzed
we have found an adequate set of integers (see tables \ref{tpgtN8} and
\ref{tpgtN16} bellow).

It results very instructive to present explicitly the first two
examples for $k=2$ and $k=3$ corresponding to $N=8$ and $N=16$, that,
according to Eq. \eqref{epinf2k}, the probabilities are

\begin{equation}
\label{elimp8}
P^{(N=8)}_\infty(t)\,=\, \frac{1}{16}
\left( 1 + \cos(2\; t) - 2 \cos(\sqrt{2} \;t) \right)^2 \,,
\end{equation}

\noindent and

\begin{equation}
\label{elimp16}
	P^{(N=16)}_\infty(t)\,=\,\frac{1}{64}
\left( 1 + \cos[2\,t] + 2 \,\cos[\sqrt{2} \;t] -
2\,\cos\left[\sqrt{2-\sqrt{2}}\;t\right]-
  2\,\cos\left[\sqrt{2 + \sqrt{2}}\;t \right] \right)^2 \,.
\end{equation}

For $N=8$, the only irrational number in Eq. \eqref{elimp8} is
$\sqrt{2}$, whose best rational approximations are given
by the Newton-Rapshon succession, that have been already shown in 
Table~\ref{new-tpgtN8}, here we tabulate the values calculated for 
$P_\infty$,

\begin{table}[h!]
\begin{center}
\begin{tabular}{c|c|c}
 $j$&  $q_j$ & $P_\infty(t_j=q_j\,\pi)$  \\ \hline
 0&  $1$ & $0.40$\\
  1&  $2$ & $0.863$\\
 2&   $12$ & $0.996$\\
 3&   $408$ & $0.999996$\\
 4&    $470832$ & $0.999999999997$\\
 5&    $627013566048$ & $0.999999999999999999999998$\\
 6 & \hspace{2mm}  $1111984844349868137938112$\hspace{2mm} & \hspace{2mm}
$0.9999999999999999999999999999999999999999999999995$ \hspace{2mm}
\end{tabular}
\end{center}
\caption{Values of the integer $q_j$ and the transmission probability in the 
strong coupling limit, $P_\infty(t_j)$, for a spin chain with length $N=8$. The 
corresponding values of $p_j$ were tabulated in Table~\ref{new-tpgtN8}. Note 
that for the last value tabulated $\varepsilon=| 1- 
P_{\infty}|< 5 \times 10^{-49}$.\label{tpgtN8}}
\end{table}

It is worth to remark that the only irrational number that it is necessary to 
approximate for the chain with $N=8$, in the strong coupling limit, is the same 
that for a homogeneous chain with $N=4$, so the asymptotic 
behavior of $t_{\varepsilon}$ for $N=8$, in the strong coupling limit, and for 
$N=4$ , for finite values of $J_2$ are the same,. Moreover, this says that if 
the dependence of $t_{\varepsilon}$ with $\varepsilon$ for $N=8$ and $J_2$ 
small is different from that of the chains with $N=4$ and $J_2$ small, then 
there must be a crossover that marks the regime change between the small $J_2$ 
behavior and the strong coupling limit.

For $N=16$ we have to use the Dirchlet's Approximation Theorem in order
to approximate three irrational numbers. The first nine approximations to
these numbers 

\begin{equation}
\label{eeaN16}
M_j=2^j \quad; \quad
\sqrt{2}\simeq\,p_j/q_j \quad ;\quad \sqrt{2-\sqrt{2}}\,\simeq\,r_j/q_j 
\quad ;\quad
\sqrt{2+\sqrt{2}}\simeq\,
s_j/q_j\quad;\quad j=1,2,\ldots \,.
\end{equation}

\noindent and the corresponding probabilities $P_\infty$ can be 
observed in Table \ref{tpgtN16}, see Appendix~\ref{ap:tables}.

\begin{figure}[hbt]
\includegraphics[width=0.7\linewidth]{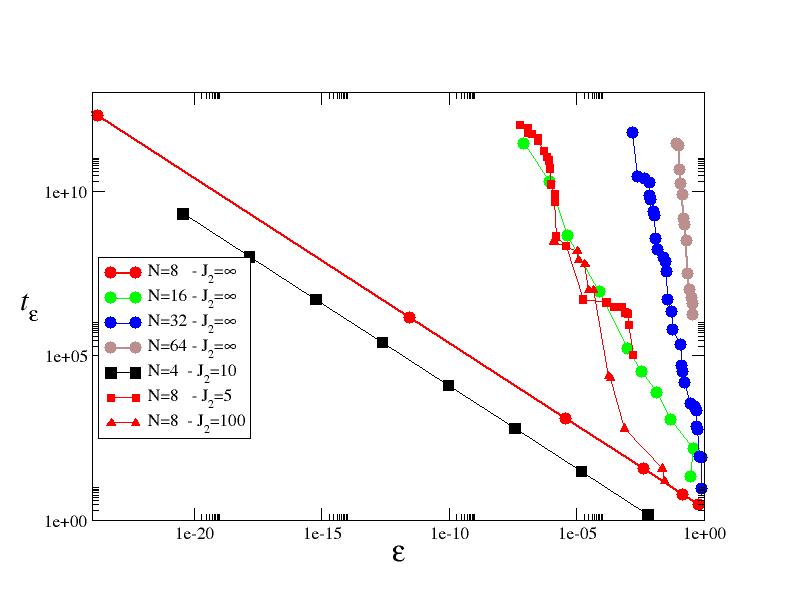}
\caption{The PGT time as a function of $\varepsilon$. The 
data in the figure was calculated for chains with lengths $N=4,8,16,32$ 
and $64$,  which are shown using black, red, green, blue and brown dots, 
respectively. The data points obtained in the strong coupling limit correspond 
to the circular dots, while the data corresponding to finite values of $J_2$ 
are shown using dots with different shapes. For instance, for $N=8$, the data 
for $J=5$ and $100$ are shown using red squares and triangles, respectively. 
The curves joining the different data points behave in the limit
($\varepsilon \longrightarrow 0$ can be fitted with power laws. See the text 
for details. \label{fig:conjetura}}
\end{figure}

At this point we can summarize our findings about the behavior of 
$t_{\varepsilon}$, in the weak and strong limit coupling, in 
Figure~\ref{fig:conjetura}.

There is a number of features that can be extracted from the Figure, note that 
the scale is log-log. First, it is clear $t_{\varepsilon}$ diverges 
asymptotically as a power law and that the power law exponent depends on the 
chain length, compare the behavior of the set of points that correspond to 
chains with $N=4$, $N=8$ that were obtained from chains with exchange 
coefficients ranging from the weak to the strong coupling limit. Second, the 
simpler expressions for the transferred population in the strong coupling limit 
allow us to obtain data for chains with length up to $N=64$, while the region 
where $J_2$ does not correspond to the strong coupling limit (SCL) can not be 
explored so  thoroughly. This is, among other traits, the more interesting 
information that the SCL offers, {\em i.e.} the asymptotic behavior of the PGT 
time in the SCL for a chain with length $2N$ is equal to the asymptotic behavior 
of the PGT time for chain with length $N$ in the ``weak'' coupling limit. 

The crossover behavior between both limits, the SCL and the one that we 
termed weak coupling limit (WCL), can be better 
observed by looking at the data and curve corresponding to a chain with $N=8$ 
and $J_2=100$. For $\varepsilon$ large enough, and shorter PGT times, the curve 
is basically parallel to the curves corresponding to $N=4$, while for smaller 
values of $\varepsilon$, and larger values of the PGT times, the curve becomes 
parallel to the ones corresponding to $N=8$, see the data shown using red 
dot  triangles pointing upwards.  

From the data shown in Figure~\ref{fig:conjetura} it is clear that, 
asymptotically, $t_{\varepsilon} \simeq 1/\varepsilon^{f(N)}$, where $f(N)$ is 
an increasing function of $N$ in both regimes, SCL and WCL. Compelling as this 
results seems we have not been able to grasp the exact form of $f(N)$ in 
despite that the numerical values of the exponents are easily obtained by 
fitting the results. The exponents, for each one of the four sets of curves, 
seem to be $1/2, 3/2, 2$ and $3$. As it will be discussed later, it is clear 
that more data could be necessary to understand these features. 

\begin{figure}[hbt]
\includegraphics[width=0.5\linewidth]{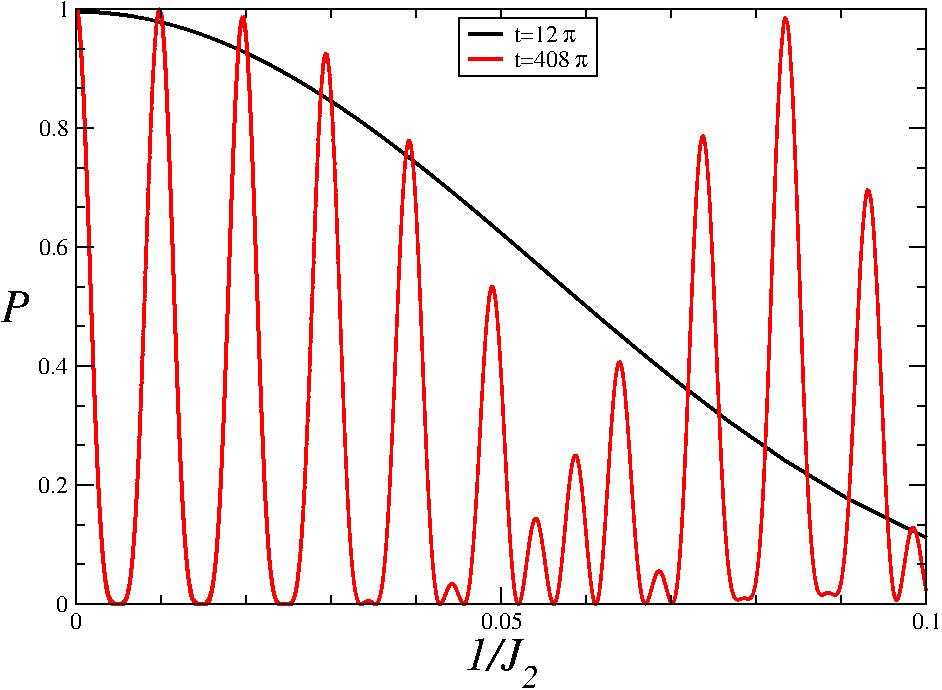}
\caption{$P$ vs. $1/J_2$ for a chain with $N=8$. The transmission probability 
is plotted for two different times, $t=12\,\pi$ y $t=408\,\pi$, while the 
coupling  $J=2\in \left[10,10000\right]$. This particular interval 
was chosen to include two times tabulated in Table~\ref{tpgtN8}, both
were calculated in the strong coupling limit.\label{fp8j2}}
\end{figure}

The crossover observed in the behavior of $t_{\varepsilon}$ can also be studied 
by changing the quantity that is kept fixed and the one 
that changes,  the PGT time and the strength of the $J_2$ coupling, 
respectively. 
In Figure~\ref{fp8j2} we show the behavior of the transmission probability 
$P(J_2,t_{\varepsilon})$ for two different values of $t_{\varepsilon}$ for a 
chain with $N=8$ spins.

The two probabilities shown in Fig.~\ref{fp8j2} correspond to $t=12\,\pi$ and
$t=408\,\pi$, which are shown as  continuous black and red curves, 
respectively. It is clear 
that for the shorter time the transmission probability reaches values close to 
the unity for $J_2>100$, but the actual value must not be precisely tuned in 
order to reach a QST with high fidelity. In contradistinction, when the PGT 
time considered is the larger one, the transmission probability shows several 
peaks that reach values close to the unity. The sharpness of the peaks 
indicates that the actual value of $J_2$ must be tuned with some care.

\section{Discussion and Conclusions}\label{sec:conclusions}

Although the results concerning the PGT time are obtained for chains with 
staggered exchange couplings, we think that the strong dependency of the PGT 
time with the chains length and the actual value of $\varepsilon$ could be  one 
of the reasons behind the difficulties that have precluded so far the proper
estimation of the PGT time in terms of $N$ and $\varepsilon$ found in previous 
works. On the other hand, the crossover observed between the SCL and the WCL is 
attributable to the staggered couplings that we have considered in this 
paper. 

In the Strong Coupling Limit the biggest challenge to analyzing the behavior of 
larger 
chains, for instance, $N=128$, is that looking for rational 
approximations of more and more irrational numbers with the same common  
denominator is computationally consuming. With a more efficient 
searching algorithm, the study of larger chains is, in principle,
doable. In the case of the WCL, the difficulty is twofold. There are more 
irrational numbers to approximate, and the exact analytical expressions are 
exceedingly convoluted with hundreds, or thousands, of terms. 

The power-law used to fit the different data sets compatible 
with PGT results in integer or semi-integer exponents. The values calculated 
are equal to 1/2, 1, 3/2, 2, and 3 for chains with N=4, 6, 8, 10, 16, 32, and 
64. The data for the last three lengths correspond to the Strong Coupling 
Limit. Remember that, as shown by our results, the PGT times obtained using the 
SCL for a chain with length $N=2^k$ scale with $\varepsilon$ as the PGT times 
obtained for chains with half this length.

For larger chains, the numerical analysis is harder to implement since 
there are data sets with times ranging over ten orders of magnitude or more.  
Interestingly enough, the quality indexes of the regression performed to obtain 
the exponents indicate that the fittings are pretty good. 

All in all, there is evidence that the exponent of the power-law is an 
increasing function of the chain size, but there is no indication of how to 
construct an analytical expression for this function. 

From the point of view of the QST, the staggered option seems very appealing
because of its simple architecture, and the fact that $t_{\varepsilon}$ at 
fixed 
values of $\varepsilon$ is a decreasing function of $J_2$. Moreover, in the SCL 
the scaling of  $t_{\varepsilon}$ corresponds to a spin chain with half 
the actual length. Even with this definite improvement concerning the 
performance of 
homogeneous chains, reaching values of the fidelity close to the 
unity, say bigger than $0.995$ could result in extremely long arrival times, 
see Figure~\ref{fig:conjetura}, even for very short chains.

The finding of chains that show PGT with lengths $N=3\times 2^k$ and $N=5\times 
2^k$ suggests that 
the regime is broadly accessible, with different interactions and architectures. 
If these chains can transfer quantum states more efficiently than the 
homogeneous 
ones deserves a deeper study. 

 The problem, of course, is that the time scale needed to observe PGT on 
homogeneous Heisenberg chains is orders of magnitude larger than the time scale 
considered in \cite{Bose2003}.  

For intermediate chain lengths ($N$ between 16 and 64), Bose analyzed the time 
dependency of the fidelity in a time window that went up to 4000.  That upper 
time is between two and ten orders of magnitude shorter than the one needed to 
pick up that the homogeneous Heisenberg chains, with lengths that are powers of 
two, effectively have pretty good transmission. 

We consider our findings of the time scale needed to achieve PGT, together with 
the expanded family of chain lengths that support this regime, as the main 
contributions of the present work.

\appendix

\section{Expressions for the cosine functions evaluated at fractions of Pi} 
\label{apcos}

From inspection and direct calculation it is easy to check that:

\begin{eqnarray}
\cos(\pi/2^2)=\frac{1}{2}\sqrt{2} \\ \nonumber
\cos(\pi/2^3)=\frac{1}{2}\sqrt{2+\sqrt{2}}  \\ \nonumber
\cos(\pi/2^4)=\frac{1}{2}\sqrt{2+\sqrt{2+\sqrt{2}}}  \\ \nonumber
\cos(\pi/2^5)=\frac{1}{2}\sqrt{2+\sqrt{2+\sqrt{2+\sqrt{2}}}}  \\ \nonumber
\end{eqnarray}

\noindent Looking these expressions we propose 

\begin{equation}
\label{ecpis2k}
\cos(\pi/2^{k+1})=\frac{1}{2} \overbrace{
\sqrt{2+\sqrt{2+\sqrt{2+\ldots +\sqrt{2}}}}}^{k \mbox{ times}} \,.
\end{equation}

\noindent and we can proof this using mathematical induction. 
We have already proved for $k=1$ and, assuming the expression Eq. 
\eqref{ecpis2k} to be true, we need to prove

\begin{equation}
\label{ecpis2kp1}
\cos(\pi/2^{k+2})=\frac{1}{2} \overbrace{
\sqrt{2+\sqrt{2+\sqrt{2+\ldots +\sqrt{2}}}}}^{k+1 \mbox{ times}} \,.
\end{equation}

\noindent with this aim we write 

\begin{equation}
\cos(\pi/2^{k+1})=\cos(2\,\pi/2^{k+2})=2\,\cos^2(\pi/2^{k+2})-1\,,
\end{equation}

\noindent using Eq. \eqref{ecpis2k} in order to replace $\cos(\pi/2^{k+1})$
we can obtain Eq. \eqref{ecpis2kp1}. Then, our initial statement is formally
proven.

We can obtain a similar expression for the sine function using 
$\cos^2(\alpha)+\sin^2(\alpha)=1$, 

\begin{equation}
\label{espis2k}
\sin(\pi/2^{k+1})=\frac{1}{2} \overbrace{
\sqrt{2-\sqrt{2+\sqrt{2+\ldots +\sqrt{2}}}}}^{k \mbox{ times}} \,,
\end{equation}

\section{Localization of Eigenvectors}
\label{ap:loc-eigenvectors}

It is interesting to point out that the delicate tuning necessary to achieve 
PGT or avert it, as was shown in Section~\ref{sec:examples} is, fundamentally, 
obtained in terms of  spectral conditions. Since the 
transmission probabilities  depend only on the first and the last eigenvectors 
components, when the eigenvectors are written on the site basis, it is natural 
to look at their behavior as function of $J_2$.

Using previous expressions we can also write the localization of
the eigenvectors in any site ($i$) for the $N=4$ Heisenberg Spin chain.
$l_\alpha^{(i)}=|\langle E_\alpha|i\rangle|^2$ as

\begin{equation}
\label{eloc4}
\begin{array}{lclclcl}
	l_1^{({1,2,3,4})}&=&\frac{1}{4}  &  & & &  \\ 
	l_2^{({1,4})} &=&C_2^2 & \qquad & l_2^{({2,3})}& = &C_4^2 \\ 
	l_3^{({1,2,3,4})}&=&\frac{1}{4} &  &  &  &\\ 
	l_4^{({1,4})}&=&C_4^2 & & l_4^{({2,3})}&=&C_2^2 \,,
\end{array}
\end{equation}

\begin{figure}[hbt]
\includegraphics[width=0.3\linewidth]{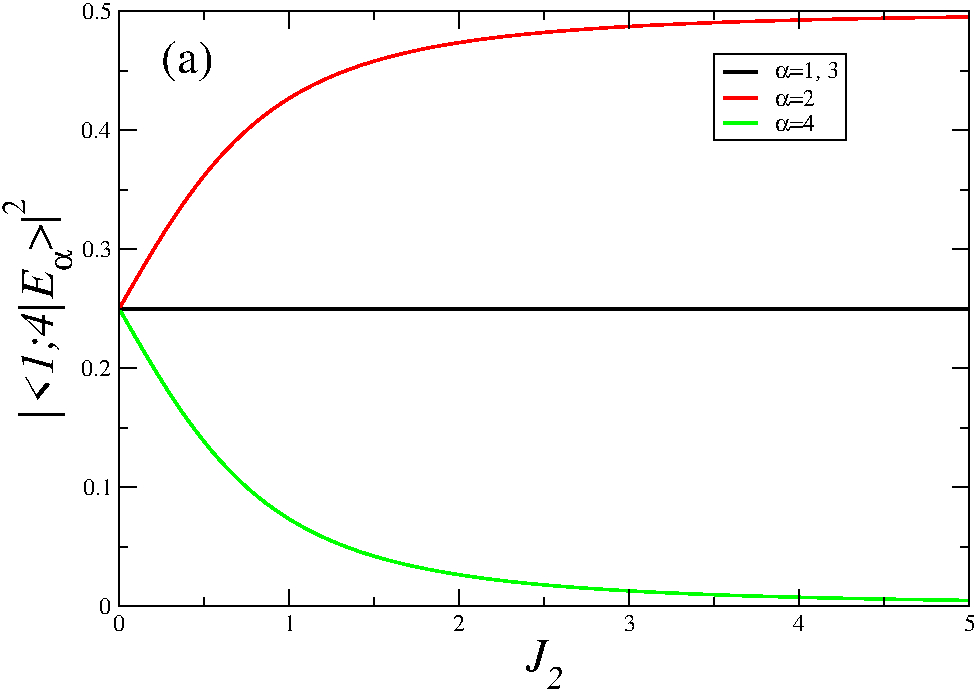} 
\includegraphics[width=0.3\linewidth]{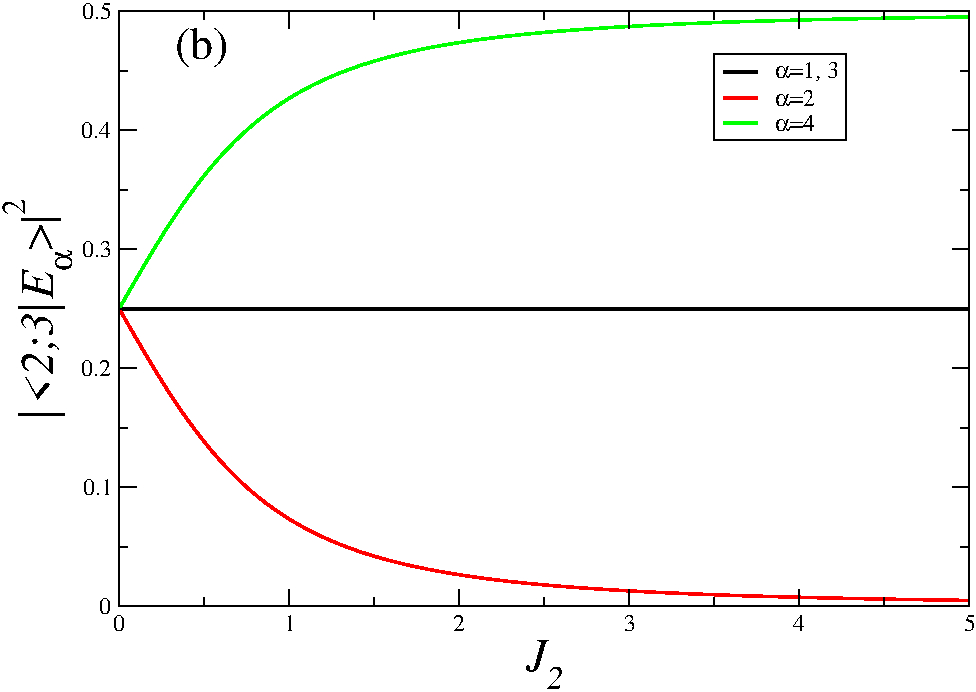}
\caption{The localization coefficients as functions of the 
coupling $J_2$ (Eq.~\ref{eloc4}) for the four 
eigenvectors of a chain with four spins.(a) For states 1 and 4
(b) For states 2 and 3.}\label{faloc4}
\end{figure}

\noindent In Fig. \ref{faloc4} we plot the localization $l_\alpha^{(i)}$

As expected, the $l_{\alpha}^{(i)}$ are smooth functions of $J_2$. This is 
in agreement  with interpretation that the behavior observed in 
Fig.~\ref{fN4J243} is owed to the fulfillment of the needed spectral 
conditions.

\begin{figure}[hbt]
\includegraphics[width=0.32\linewidth]{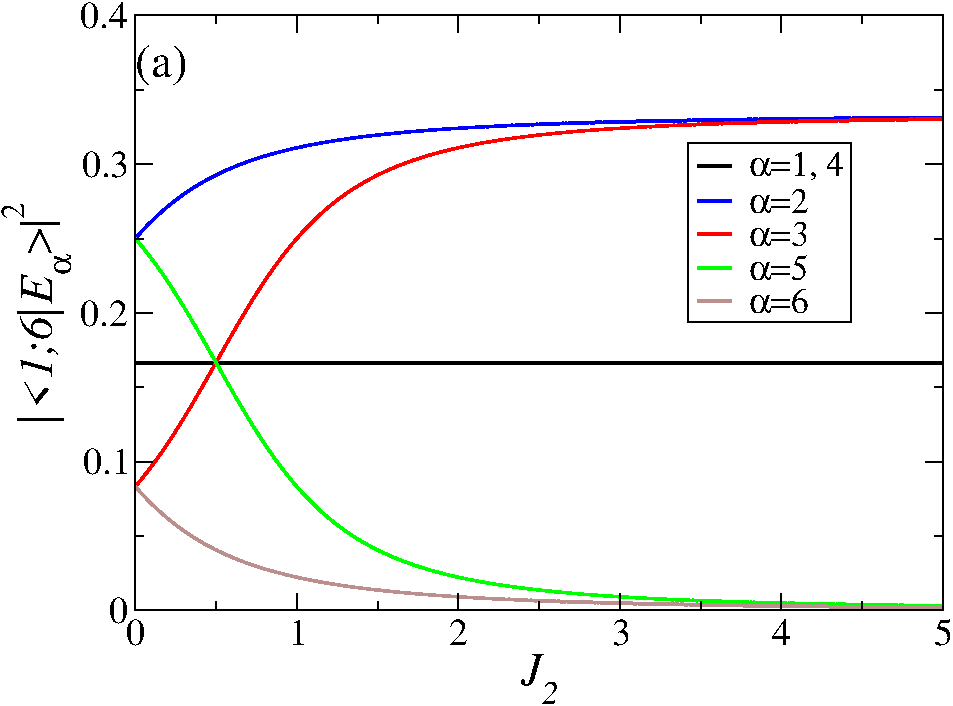} \hspace{0.1cm}
\includegraphics[width=0.32\linewidth]{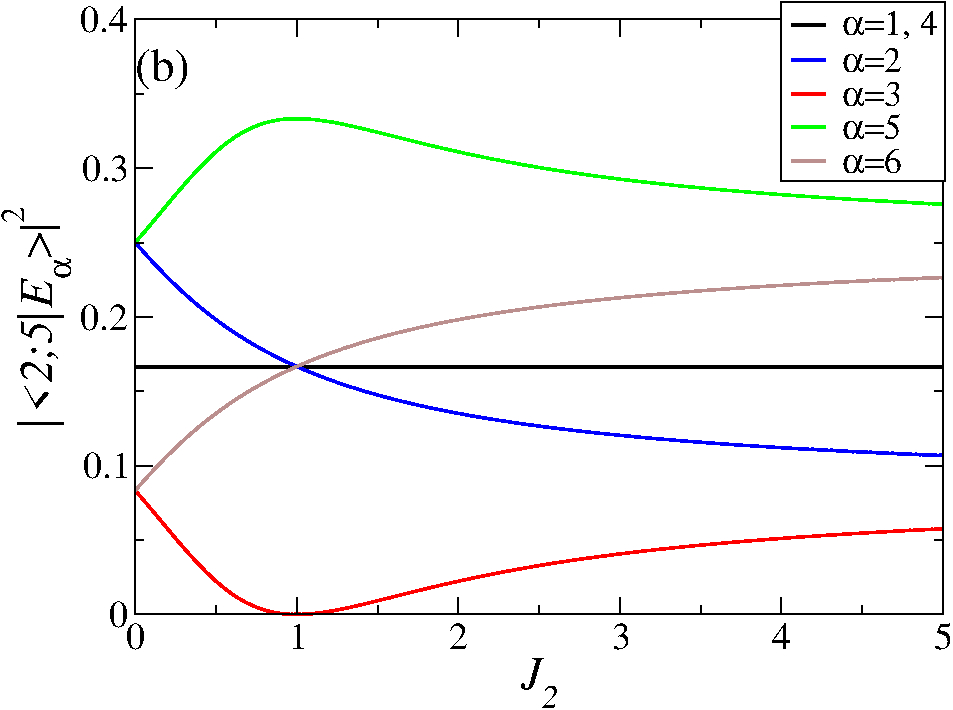} \hspace{0.1cm}
	\includegraphics[width=0.32\linewidth]{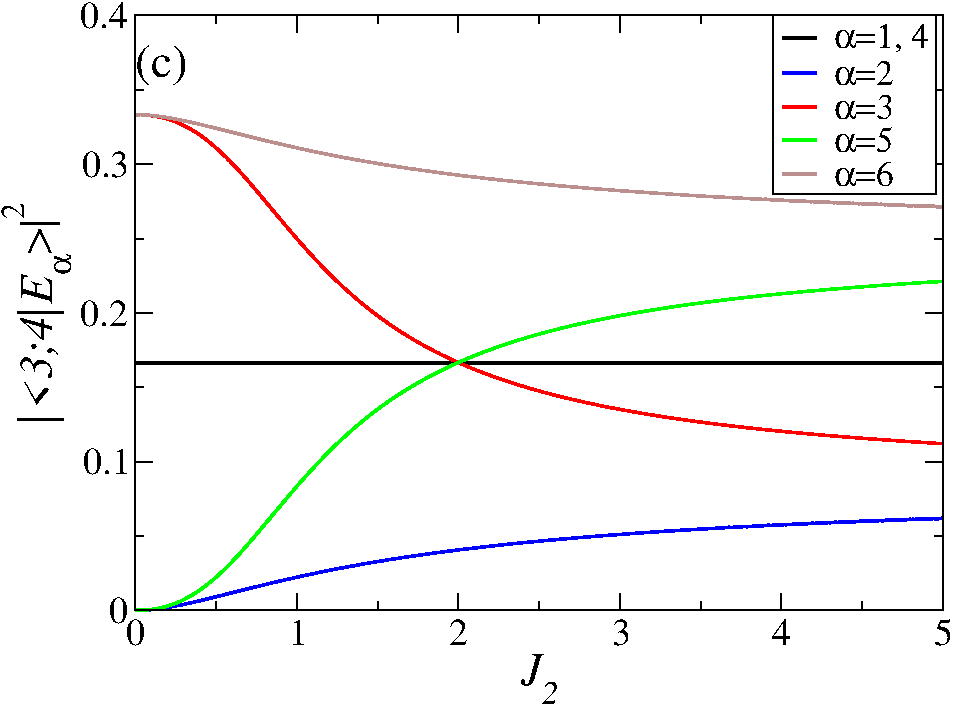} \vspace{0.1cm}
\caption{The localization coefficients as functions of the 
coupling $J_2$ (Eq.~\ref{eloc4}) for the four 
eigenvectors of a chain with six spins.(a) For states 1 and 6.
(b) For states 2 and 5.
(c) For states 3 and 4.
	\label{faloc6}}
\end{figure}

For the kind of spin chains that we considered in this work the localization 
probabilities in the different sites of the chain behave in the same way. For 
lengths large enough, there is reduced number of eigenvectors that become 
``localized`` at the extremes of the chain. For this eigenvectors, the 
localization $l_{\alpha}^{(i)}$ are increasing functions of $J_2$, see the red 
and blue curves in Fig. \ref{faloc6} a). The other eigenvectors are 
''extended`` 
along the chain, the functions  $l_{\alpha}^{(i)}$ of those eigenvectors become 
negligible at the extremes of the chain. For the extended eigenvectors, the 
localization in the interior of the chain are not decreasing or increasing 
functions of $J_2$, but their value slowly goes to the value $1/N$, see 
Fig.~\ref{faloc6} b) and c), where the behavior described can be observed even 
when the data correspond to a spin chain with $N=6$. Note that the eigenvector 
that is perfectly distributed along the length of the chain is depicted with 
the black curve, in both Fig.~\ref{faloc4} and \ref{faloc6}.

\section{Tables}
\label{ap:tables}

\begin{table}[h!]
\begin{center}
\begin{tabular}{c|c|c|c}
 $j$& $p_j$&  $q_j$ & $P(J_2=1;t_j=q_j\,\pi/2)$  \\ \hline
1 & 1 & 1& 0.5170 \\
 2&  3&  $2$ & $0.8962$\\
  3&  7& $5$ & $0.9815$\\
 4 & 17 & 12 & 0.9968 \\
 5& 41&   $29$ & $0.99945$\\
 6 & 99 & 70 & 0.999906 \\
 7& 239&   $169$ & $0.9999838$\\
 8 & 577 & 408   &   0.9999972
\end{tabular}
\end{center}
\caption{Values of $p_j,\;q_j$ y $P(1;t_j)$ for a chain with $N=4$ spins.
The index $j$ gives the order of the continued fraction approximation of 
$\sqrt{2}$ as $p_j/q_j$.\label{tN4J21}}
\end{table}

\begin{table}[h!]
\begin{center}
\begin{tabular}{c|c|c|c}
 $j$& $p_j$&  $q_j$ & $P(J_2=2;t_j=q_j\,\pi/2)$  \\ \hline
 1&  2&  $1$ & $0.8459$\\
 2& 9 & 4 & 0.9908 \\
  3&  38& $17$ & $0.99949$\\
 4& 161&   $72$ & $0.999971$\\
 5& 682&   $305$ & $0.9999984$\\
 6&2889 & 1292   & 0.999999911\\
 7& 12238  & 5473  & 0.9999999951 \\
 8& 51841 &  23184 & 0.99999999972
\end{tabular}
\end{center}
\caption{Values of $p_j,\;q_j$ y $P(2;t_j)$ for a chain with $N=4$ spins.
The index $j$ gives the order of the continued fraction approximation of
$\sqrt{5}$ as $p_j/q_j$.\label{tN4J22}}
\end{table}

\begin{table}[h!]
\begin{center}
\begin{tabular}{c|c|c|c}
 $j$& $p_j$&  $q_j$ & $P(J_2=5;t_j=q_j\,\pi/2)$  \\ \hline
 1& 5  &  $1$ & $0.4981$\\
 2& 51 & 10 & 0.99975 \\
  3& 515  & 101 & 0.9999976 \\
 4& 5201&  1020 & 0.999999976 \\
 5& 52525 & 10301   & 0.99999999977\\
 6& 530451 & 104030   & 0.9999999999977\\
 7& 5357035  & 1050601  & 0.999999999999978  \\
 8& 54100801 &  10610040 & 0.99999999999999978
\end{tabular}
\end{center}
\caption{Values of $p_j,\;q_j$ y $P(5;t_j)$ for a chain with $N=4$ spins.
The index $j$ gives the order of the continued fraction approximation of
$\sqrt{26}$ as $p_j/q_j$.\label{tN4J25}}
\end{table}

\begin{table}[h!]
\begin{center}
\begin{tabular}{c|c|c|c}
 $j$& $p_j$&  $q_j$ & $P(J_2=10;t_j=q_j\,\pi/2)$  \\ \hline
 1& 10  &  $1$ & 0.4956\\
 2& 201 & 20 &0.999985 \\
  3& 4030 & 401 & 0.999999962 \\
 4& 80801 & 8040 & 0.999999999905 \\
 5& 1620050 & 161201 & 0.99999999999976\\
 6& 32481801 & 3232060 & 0.99999999999999941 \\
 7& 651256070 & 64802401 & 0.9999999999999999985 \\
 8& 13057603201 & 1299280080 & 0.9999999999999999999963
\end{tabular}
\end{center}
\caption{Values of $p_j,\;q_j$ y $P(10;t_j)$ for a chain with $N=4$ spins.
The index $j$ gives the order of the continued fraction approximation of
$\sqrt{101}$ as $p_j/q_j$.\label{tN4J210}}
\end{table}

\begin{table}[hbt]
\begin{center}
\begin{tabular}{c|c|c|c|c|c}
 $j$& $q_j$&  $p_j$ &$r_j$&$s_j$& $P_\infty(t_j=q_j\,\pi)$  \\ \hline
  1 & 7 & 10&  5 & 11 & 0.7065 \\
  2 & 48 & 68 & 36 &87 & 0.6319 \\
  3& 362 & $512$ & 277  & 669  &0.9533 \\
 4 &  2489  &3520   &1905 & 4588 & 0.9860\\
 5 & 10608 & 15002 & 8119  & 19601 & 0.9965\\
6 &227291 & 321438  &173961 & 419955 & 0.9994\\
7  & 1562648 & 2209918 & 1195999 & 2887397 &  0.9998 \\
8 &  2758647 & 3901316 &  2111302  &5097315 & 0.99992 \\
9 & 28384580 & 40141858 & 21724617 & 52447865 & 0.99998 \\
\end{tabular}
\end{center}
\caption{$N=16$. The first nine quartets of values $p_j,\;q_j,\;r_j,\;s_j$ and 
$P_\infty(t_j)$ that satisfy the parity condition. The 
ratios $p_j/q_j$, $r_j/q_j$ and $s_j/q_j$ are successive approximations for 
the irrational numbers $\sqrt{2}, \sqrt{2-\sqrt{2}}$ and $\sqrt{2+\sqrt{2}} $, 
respectively
 \label{tpgtN16}}
\end{table}

\end{document}